\def \inparg{\leftskip = 40pt\rightskip = 40pt}
\def \outparg{\leftskip = 0 pt\rightskip = 0pt}

\def\npb{{Nucl.\ Phys.\ }{\bf B}}
\def\plb{{Phys.\ Lett.\ }{ \bf B}}

\def\prd{{Phys.\ Rev.\ }{\bf D}}

\def\prl{Phys.\ Rev.\ Lett.\ }

\def\ov #1{\overline{#1}}
\def\wt #1{\widetilde{#1}}
\def\Tr{\mathop{\rm Tr}}

\def\th{{\theta}}
\def\bth{{\overline{\theta}}}

\def\frak#1#2{{\textstyle{{#1}\over{#2}}}}

\def\ntwo{${\cal N} = 2$}

\def\Ahat{\hat A}

\def\frak#1#2{{\textstyle{{#1}\over{#2}}}}

\def\lambdabar{\bar\lambda}

\def\Dtil{\tilde D}

\def\Ftil{\tilde F}
\def\Fhat{\hat F}

\def\sigmabar{\bar\sigma}
\def\phibar{\bar\phi}

\def\psibar{\bar\psi}
\def\Fbar{\bar F}

\def\Dtil{\tilde D}

\def\Ncal{{\cal N}}
\def\Ftil{\tilde F}
\def\alphadot{\dot\alpha}
\def\betadot{\dot\beta}

\def\pa{\partial}

\input harvmac
\input epsf
% +--------------------------------------------------------------------+
% |                                                                    |
% |                           TABLES.TEX                               |
% |                                                                    |
% |                     Ray F. Cowan  15-Feb-85                        |
% |                                                                    |
% |                       Princeton University                         |
% |                                                                    |
% |          Present Address:  Laboratory for Nuclear Science          |
% |                            M.I.T.                                  |
% |                            Cambridge, MA 02139                     |
% |                                                                    |
% |                   E-mail:  rfc@slacvm.slac.stanford.edu            |
% |                                                                    |
% |                                                                    |
% |                     Last Revision: 17-Apr-86                       |
% |                                                                    |
% |   Macros I find handy for making tables.  See TABLEDOC TEX for     |
% |   a longer description.  The token-counting macros are straight    |
% |   from the TeXbook's "Dirty Tricks" appendix.                      |
% |                                                                    |
% +--------------------------------------------------------------------+
%
\newbox\hdbox%
\newcount\hdrows%
\newcount\multispancount%
\newcount\ncase%
\newcount\ncols% This is the number of primary text columns in the table.
\newcount\nrows%
\newcount\nspan%
\newcount\ntemp%
\newdimen\hdsize%
\newdimen\newhdsize%
\newdimen\parasize%
\newdimen\spreadwidth%
\newdimen\thicksize%
\newdimen\thinsize%
\newdimen\tablewidth%
\newif\ifcentertables%
\newif\ifendsize%
\newif\iffirstrow%
\newif\iftableinfo%
\newtoks\dbt%
\newtoks\hdtks%
\newtoks\savetks%
\newtoks\tableLETtokens%
\newtoks\tabletokens%
\newtoks\widthspec%
%
%  Book-keeping stuff--see how often these macros are called.
%
%  MOD RFC 900221.
%  Removed usage logging:  it's too complicated under VM/XA.
%\immediate\write15{%
%CP SMSG GJMSINK TEXTABLE --> TABLE MACROS V. 851121 JOB = \jobname%
%}%
%
%  Turn on table diagnostics.
%
\tableinfotrue%
\catcode`\@=11%  Allows use of "@" in macro names, like PLAIN.TEX does.
%  Debugging aid.  Writes #1 on the
%                                    user's terminal and in the log file.
%
%  Define the \tstrut height, depth in terms of the x_height parameter.
%
\def\tstrut{\vrule height3.1ex depth1.2ex width0pt}%
\def\and{\char`\&}%  Allows us to get an `&' in the text.  This is the
%                    same as using the PLAIN TeX macro \&.
\def\tablerule{\noalign{\hrule height\thinsize depth0pt}}%
\thicksize=1.5pt%  Default thickness for fat rules.  The user should feel
%                  free to change this to his preference.
\thinsize=0.6pt%   Default thickness for thin rules.
\def\thickrule{\noalign{\hrule height\thicksize depth0pt}}%
\def\ctr#1{\hfil\ #1\hfil}%
%
%
%
%  Here are things for controlling the width of the finished table.
%
\tablewidth=-\maxdimen%
\spreadwidth=-\maxdimen%
\def\tabskipglue{0pt plus 1fil minus 1fil}%
%
%  Stuff for centering or not.
%
\centertablestrue%
%
%
%
%  \vctr vertically centers its argument in the row.
%
\parasize=4in%
\gdef\ARGS{########}%  Produces the correct number of #'s in the preamble
%                      by the time eveything is expanded and \halign sees
%                      it.
\gdef\headerARGS{####}%  Same as \ARGS, but used in \header macros.
\def\@mpersand{&}%  Allows us to get alignment tab characters later
%                   when we have made the character "&" an active macro.
{\catcode`\|=13%  Make |'s locally active.
\gdef\letbarzero{\let|0}%  Globally define a macro that allows us to
%                          keep active |'s from being expanded in edef's.
\gdef\letbartab{\def|{&&}}%
\gdef\letvbbar{\let\vb|}%
%  This \def will cause active |'s read by
%                            \ruledtable to be converted into double
%                            alignment tabs.
}%  End of locally active |'s.
{\catcode`\&=4%  Make these alignment tabs.
\def\ampskip{&\omit\hfil&}%  This local macro skips a vertical rule.
\catcode`\&=13%  Now make &'s into active macros.
\let&0%  This allows us to expand \ampskip in the next \xdef without
%        attempting to expand the & and getting an "undefined control
%        sequence" error.
\xdef\letampskip{\def&{\ampskip}}%
\gdef\letnovbamp{\let\novb&\let\tab&}
%  This will cause active &'s read by
%                                   \ruledtable to be converted into
%                                   double tabs and an \omit'ted \vrule.
}%  End of locally active &'s.
\def\begintable{%  Here we make |'s and &'s active characters so we can
%                  interpret them as macros.  Note that this action is
%                  true only until we encounter the matching \endgroup
%                  token later at the end of the \ruledtable macro.
   \begingroup%
   \catcode`\|=13\letbartab\letvbbar%
   \catcode`\&=13\letampskip\letnovbamp%
   \def\multispan##1{%  We must redefine \multispan to count the number
%                       of primary columns, not physical columns.
      \omit \mscount##1%
      \multiply\mscount\tw@\advance\mscount\m@ne%
      \loop\ifnum\mscount>\@ne \sp@n\repeat%
   }%  End of \multispan macro.
   \def\|{%
      &\omit\widevline&%
   }%
   \ruledtable%  Now we call \ruledtable to do the real work.
}%  End of \begintable macro.
\long\def\ruledtable#1\endtable{%
%
%  This macro reads in the user's data entries
%  and converts them into a ruled table.
%
%  Important note:  Many macros and parameters are re-defined here, and
%  these must be kept local to the table macros to avoid conflict with
%  their use outside of tables.  This is done by the \begingroup token
%  macro \begintable and the \endgroup token at the end of
%  this macro.
%
   \offinterlineskip%  Needed to make rules touch each other.
   \tabskip 0pt%  Needed for same reason as \offinterlineskip.
   \def\widevline{\vrule width\thicksize}%  Make outer \vrule's wider.
   \def\endrow{\@mpersand\omit\hfil\crnorm\@mpersand}%
   \def\crthick{\@mpersand\crnorm\thickrule\@mpersand}%
   \def\crthickneg##1{\@mpersand\crnorm\thickrule
          \noalign{{\skip0=##1\vskip-\skip0}}\@mpersand}%
   \def\crnorule{\@mpersand\crnorm\@mpersand}%
   \def\crnoruleneg##1{\@mpersand\crnorm
          \noalign{{\skip0=##1\vskip-\skip0}}\@mpersand}%
   \let\nr=\crnorule%  A shorter abbreviation.
   \def\endtable{\@mpersand\crnorm\thickrule}%
   \let\crnorm=\cr%  Allows us to use \cr for our own purposes.
%
%  Cause user-typed \cr's to follow a row with a \tablerule.
%
   \edef\cr{\@mpersand\crnorm\tablerule\@mpersand}%
   \def\crneg##1{\@mpersand\crnorm\tablerule
          \noalign{{\skip0=##1\vskip-\skip0}}\@mpersand}%
   \let\ctneg=\crthickneg
   \let\nrneg=\crnoruleneg
   \the\tableLETtokens%  Get the user's extra \let's, if any.
%
%  Put the data entries into a token register so we can scan through them
%  and see what the user is asking us to do.
%
   \tabletokens={&#1}%  We add an extra alignment tab to the beginning
%                       of the first row to allow for the first \vrule.
%
%  Now count how many rows are in the table and return the result in
%  count register \nrows; do the same for columns, and return that
%  in register \ncols.
%
   \countROWS\tabletokens\into\nrows%
   \countCOLS\tabletokens\into\ncols%
%
%  Now do a little arithmetic to convert the number of primary columns
%  into the number of physical columns that the alignment preamble must
%  prepare for;  similarly for rows.
%
   \advance\ncols by -1%
   \divide\ncols by 2%
   \advance\nrows by 1%
%
%  Tell the user how many rows and columns we found in his data, if he
%  wants to know.
%
   \iftableinfo %
      \immediate\write16{[Nrows=\the\nrows, Ncols=\the\ncols]}%
   \fi%
%
%  Now we actually go ahead and produce the table.
%
   \ifcentertables
      \ifhmode \par\fi%  Make sure we are in vertical mode.
      \line{%  The final table comes out as an \hbox of width the \hsize.
      \hss%  The final table will be centered left-to-right.
   \else %
      \hbox{%
   \fi
      \vbox{%
         \makePREAMBLE{\the\ncols}%  Generate the preamble.
         \edef\next{\preamble}%  This line and the next line force the
         \let\preamble=\next%    expansion of all \ARGS tokens into the
%                                appropriate number of #'s.
         \makeTABLE{\preamble}{\tabletokens}%  Go do the \halign here.
      }%  End of \vbox.
      \ifcentertables \hss}\else }\fi%  Finish the centering effect.
%                                       It is important that no spaces
%                                       follow the two `}' here.
%  }%  End of \line.
   \endgroup%  Return all local macros and parameters to their outside
%              values.
   \tablewidth=-\maxdimen%  Reset \tablewidth to normal.
   \spreadwidth=-\maxdimen% Same for \spreadwidth.
}%  End of macro \ruledtable.
\def\makeTABLE#1#2{%  Does an \halign for the \ruledtable macro.
   {%  Start of local parameter values.
   \let\ifmath0%     These macros would cause trouble if they were to be
   \let\header0%     expanded in the following \xdef; we \let them be
   \let\multispan0%  equal to a digit, because digits can't be expanded.
%
%  Set up the width specification here.
%
   \ncase=0%
   \ifdim\tablewidth>-\maxdimen \ncase=1\fi%
   \ifdim\spreadwidth>-\maxdimen \ncase=2\fi%
   \relax%  This \relax is absolutely necessary, without it the following
%           \ifcase will always take \ncase=0.
%
   \ifcase\ncase %
      \widthspec={}%
   \or %
      \widthspec=\expandafter{\expandafter t\expandafter o%
                 \the\tablewidth}%
   \else %
      \widthspec=\expandafter{\expandafter s\expandafter p\expandafter r%
                 \expandafter e\expandafter a\expandafter d%
                 \the\spreadwidth}%
   \fi %
%\out{Widthspec=[\the\widthspec]}%
%\out{Preamble=[\preamble]}%
   \xdef\next{%  We must force the preamble to be expanded BEFORE the
      \halign\the\widthspec{%
%        \halign is done;  this \edef\next{...}\next construction
%                does the trick.
      #1%  This is the preamble text.
      \noalign{\hrule height\thicksize depth0pt}%  Makes the top \hrule.
      \the#2\endtable%  This is the main body.
%
%     \noalign{\hrule height0.7pt depth0pt}%  Makes the last \hrule.
      }%  End of \halign.
   }%  End of \next.
   }%  End of local values.
   \next%  This \next must be outside of the local values, because now
%          we want those troublesome macros in the \let's above to have
%          their normal actions.
}%  End of macro \makeTABLE.
\def\makePREAMBLE#1{%  This macro generates the necessary preamble for a
%                      ruled table with #1 primary columns.
%                      (Primary columns means the number of columns NOT
%                       counting those used for vertical rules.)
   \ncols=#1%  Get the number of columns desired.
   \begingroup%  Start local parameter definitions.
   \let\ARGS=0%  This is the key to the whole thing; it prevents \ARGS
%                from being expanded in the following \edef's.
   \edef\xtp{\widevline\ARGS\tabskip\tabskipglue%
   &\ctr{\ARGS}\tstrut}%  A 1-column preamble.  Gets the sizing right.
   \advance\ncols by -1%  One column has been generated; decrement the
%                         counter.
   \loop%  Append as many further columns as needed to the preamble.
      \ifnum\ncols>0 %
      \advance\ncols by -1%
      \edef\xtp{\xtp&\vrule width\thinsize\ARGS&\ctr{\ARGS}}%
   \repeat
   \xdef\preamble{\xtp&\widevline\ARGS\tabskip0pt%
   \crnorm}%  Adds the last \vrule.
   \endgroup%  End of local parameters.
}%  End of macro \makePREAMBLE.
\def\countROWS#1\into#2{%  This counts the number of rows in #1 by
%                          looking for control sequences that end a row,
%                          e.g., \cr, \crthick, etc., and puts the result
%                          into count register #2.
   \let\countREGISTER=#2%
   \countREGISTER=0%
%  \out{In countROWS:  tokens are [\the#1]}%
   \expandafter\ROWcount\the#1\endcount%
}%
\def\ROWcount{%
   \afterassignment\subROWcount\let\next= %
}%
\def\subROWcount{%
%  \out{In subROWcount:  next is [\meaning\next]}%  Debugging aid.
   \ifx\next\endcount %
      \let\next=\relax%
   \else%
      \ncase=0%
      \ifx\next\cr %
         \global\advance\countREGISTER by 1%
         \ncase=0%
      \fi%
      \ifx\next\endrow %
         \global\advance\countREGISTER by 1%
         \ncase=0%
      \fi%
      \ifx\next\crthick %
         \global\advance\countREGISTER by 1%
         \ncase=0%
      \fi%
      \ifx\next\crnorule %
         \global\advance\countREGISTER by 1%
         \ncase=0%
      \fi%
      \ifx\next\crthickneg %
         \global\advance\countREGISTER by 1%
         \ncase=0%
      \fi%
      \ifx\next\crnoruleneg %
         \global\advance\countREGISTER by 1%
         \ncase=0%
      \fi%
      \ifx\next\crneg %
         \global\advance\countREGISTER by 1%
         \ncase=0%
      \fi%
      \ifx\next\header %
%     \out{In subROWcount:  next=header, ncase set=1}%
         \ncase=1%
      \fi%
%     \out{In subROWcount:  ncase is [\the\ncase]}%
      \relax%
      \ifcase\ncase %
         \let\next\ROWcount%
%        \out{subROWcount---> ncase=\the\ncase}%
      \or %
         \let\next\argROWskip%
%        \out{subROWcount---> ncase=\the\ncase}%
      \else %
      \fi%
   \fi%
%  \out{subROWcount---> NEXT=\meaning\next}%
   \next%
}%  End of macro \subROWcount.
\def\counthdROWS#1\into#2{%
\dvr{10}%
   \let\countREGISTER=#2%
   \countREGISTER=0%
\dvr{11}%
%  \out{In counthdROWS:  tokens are [\the#1]}%
\dvr{13}%
   \expandafter\hdROWcount\the#1\endcount%
\dvr{12}%
}%
\def\hdROWcount{%
   \afterassignment\subhdROWcount\let\next= %
}%
\def\subhdROWcount{%
%\out{In subhdROWcount:  next is [\meaning\next]}%
   \ifx\next\endcount %
      \let\next=\relax%
   \else%
      \ncase=0%
      \ifx\next\cr %
         \global\advance\countREGISTER by 1%
         \ncase=0%
      \fi%
      \ifx\next\endrow %
         \global\advance\countREGISTER by 1%
         \ncase=0%
      \fi%
      \ifx\next\crthick %
         \global\advance\countREGISTER by 1%
         \ncase=0%
      \fi%
      \ifx\next\crnorule %
         \global\advance\countREGISTER by 1%
         \ncase=0%
      \fi%
      \ifx\next\header %
%\out{In subhdROWcount:  next=header, ncase set=1}%
         \ncase=1%
      \fi%
%\out{In subhdROWcount:  ncase is [\the\ncase]}%
\relax%
      \ifcase\ncase %
         \let\next\hdROWcount%
%\out{subhdROWcount---> ncase=\the\ncase}%
      \or%
         \let\next\arghdROWskip%
%\out{subhdROWcount---> ncase=\the\ncase}%
      \else %
      \fi%
   \fi%
%\out{subhdROWcount---> NEXT=\meaning\next}%
   \next%
}%
{\catcode`\|=13\letbartab
\gdef\countCOLS#1\into#2{%
%  \out{In countCOLS:  tokens are [\the#1]}
   \let\countREGISTER=#2%
   \global\countREGISTER=0%
   \global\multispancount=0%
   \global\firstrowtrue
   \expandafter\COLcount\the#1\endcount%
   \global\advance\countREGISTER by 3%
   \global\advance\countREGISTER by -\multispancount
%  \out{countCOLS-->[\the\countREGISTER]}
}%
\gdef\COLcount{%
   \afterassignment\subCOLcount\let\next= %
}%
{\catcode`\&=13%
\gdef\subCOLcount{%
%\out{In subCOLcount: next is [\meaning\next]}
   \ifx\next\endcount %
      \let\next=\relax%
   \else%
      \ncase=0%
      \iffirstrow
         \ifx\next& %
            \global\advance\countREGISTER by 2%
            \ncase=0%
         \fi%
         \ifx\next\span %
            \global\advance\countREGISTER by 1%
            \ncase=0%
         \fi%
         \ifx\next| %
            \global\advance\countREGISTER by 2%
            \ncase=0%
         \fi
         \ifx\next\|
            \global\advance\countREGISTER by 2%
            \ncase=0%
         \fi
         \ifx\next\multispan
            \ncase=1%
            \global\advance\multispancount by 1%
         \fi
         \ifx\next\header
            \ncase=2%
         \fi
         \ifx\next\cr       \global\firstrowfalse \fi
         \ifx\next\endrow   \global\firstrowfalse \fi
         \ifx\next\crthick  \global\firstrowfalse \fi
         \ifx\next\crnorule \global\firstrowfalse \fi
         \ifx\next\crnoruleneg \global\firstrowfalse \fi
         \ifx\next\crthickneg  \global\firstrowfalse \fi
         \ifx\next\crneg       \global\firstrowfalse \fi
      \fi%  End of \iffirstrow.
\relax%\out{subCOL-->  ncase=[\the\ncase]}
% \out{subCOL-->  next=\meaning\next}
      \ifcase\ncase %
         \let\next\COLcount%
      \or %
         \let\next\spancount%
      \or %
         \let\next\argCOLskip%
      \else %
      \fi %
   \fi%
%  \out{subCOL-->  countREGISTER=[\the\countREGISTER]}
   \next%
}%
\gdef\argROWskip#1{%
%  Deletes the next balanced, undelimited argument from a
%                 token list.
% \out{---> Entering argROWskip <---}
% \out{In argROWskip:  deleted arg is [#1]}%
   \let\next\ROWcount \next%
}%  End of macro \argskip.
\gdef\arghdROWskip#1{%
%  Deletes the next balanced, undelimited argument from a
%                 token list.
% \out{---> Entering arghdROWskip <---}
% \out{In arghdROWskip:  deleted arg is [#1]}%
   \let\next\ROWcount \next%
}%  End of macro \arghdROWskip.
\gdef\argCOLskip#1{%
%  Deletes the next balanced, undelimited argument from a
%                 token list.
% \out{---> Entering argCOLskip <---}
% \out{In argCOLskip:  deleted arg is [#1]}%
   \let\next\COLcount \next%
}%  End of macro \argskip.
}%  End of active &'s.
}%  End of active |'s.
\def\spancount#1{%\out{spancount--->\meaning#1}
   \nspan=#1\multiply\nspan by 2\advance\nspan by -1%
   \global\advance \countREGISTER by \nspan
%  \out{number spancount--->\the\nspan; \the\countREGISTER}
   \let\next\COLcount \next}%
\def\dvr#1{\relax}%
% \omit\hfil%
% \parindent=0pt\hsize=1.1in\valign{%
% \vfil#\vfil&\vfil#\vfil\cr\hfil\hbox{\ Added to\ }\hfil&%
% \hfil\hbox{\ empty events\ }\hfil\cr}\hfil%
\def\header#1{%
\dvr{1}{\let\cr=\@mpersand%
\hdtks={#1}%
%\out{In header:  hdtks=[\the\hdtks]}%
\counthdROWS\hdtks\into\hdrows%
\advance\hdrows by 1%
\ifnum\hdrows=0 \hdrows=1 \fi%
%\out{In header:  Nhdrows=[\the\hdrows]}%
\dvr{5}\makehdPREAMBLE{\the\hdrows}%
%\out{In header:  headerpreamble=[\headerpreamble]}%
\dvr{6}\getHDdimen{#1}%
%\out{In header:  hdsize=[\the\hdsize]}%
%\striplastCR{#1}%
{\parindent=0pt\hsize=\hdsize{\let\ifmath0%
\xdef\next{\valign{\headerpreamble #1\crnorm}}}\dvr{7}\next\dvr{8}%
}%
}\dvr{2}}%  End of macro \header.
\def\makehdPREAMBLE#1{%This macro generates the necessary preamble for a
\dvr{3}%
%                      ruled table with \ncols primary columns.
%                      (Primary columns means the number of columns NOT
%                       counting those used for vertical rules.
\hdrows=#1%  Get the number of columns desired.
{%  Start local parameter definitions.
\let\headerARGS=0%
%  This is the key to the whole thing; it prevents \ARGS
\let\cr=\crnorm%
%                from being expanded in the followin \edef's.
\edef\xtp{\vfil\hfil\hbox{\headerARGS}\hfil\vfil}%
\advance\hdrows by -1%  One row has been generated; decrement the
%                         counter.
\loop%  Append as many further rows as needed to the preamble.
\ifnum\hdrows>0%
\advance\hdrows by -1%
\edef\xtp{\xtp&\vfil\hfil\hbox{\headerARGS}\hfil\vfil}%
\repeat%
\xdef\headerpreamble{\xtp\crcr}%
}%  End of local parameters.
\dvr{4}}%  End of \makehdPREAMBLE.
\def\getHDdimen#1{%
%\out{In getHDdimen:  Arg 1=[#1]}%
\hdsize=0pt%
\getsize#1\cr\end\cr%
}%  End of macro getHDdimen.
\def\getsize#1\cr{%
%\out{In getsize:  Arg 1=[#1]}%
%  Here we have to check arg#1 and see if the first token in #1 is an
%    \end; if so, we stop, else we check the width of arg#1.
%  We recall that each arg#1 will be terminated with a \cr token.
\endsizefalse\savetks={#1}%
%\out{In getsize:  the savetks = [\the\savetks]}%
\expandafter\lookend\the\savetks\cr%
%\out{In getsize:  ifendsize = [\meaning\ifendsize]}%
\relax \ifendsize \let\next\relax \else%
\setbox\hdbox=\hbox{#1}\newhdsize=1.0\wd\hdbox%
\ifdim\newhdsize>\hdsize \hdsize=\newhdsize \fi%
%\out{In getsize:  hdsize=[\the\hdsize]}%
%\out{In getsize:  newhdsize=[\the\newhdsize]}%
\let\next\getsize \fi%
\next%
}%
\def\lookend{\afterassignment\sublookend\let\looknext= }%
\def\sublookend{\relax%
%\out{In sublookend:  looknext = [\looknext]}%
\ifx\looknext\cr %
%\out{In sublooknext:  looknext=cr}%
\let\looknext\relax \else %
%\out{In sublooknext:  looknext/=cr}%
   \relax
   \ifx\looknext\end \global\endsizetrue \fi%
   \let\looknext=\lookend%
    \fi \looknext%
}%
%
%  Allow the user to make his own names for crthick, etc.
%
\def\tablelet#1{%
   \tableLETtokens=\expandafter{\the\tableLETtokens #1}%
}%
\catcode`\@=12%  Change @'s back to their normal category code.

{\nopagenumbers
\line{\hfil LTH 709}
\line{\hfil hep-th/0607195}
%\line{\hfil Revised Version}
\vskip .5in
\centerline{\titlefont One-loop renormalisation of }
\centerline{\titlefont $\Ncal=\frak12$ supersymmetric gauge theory}
\centerline{\titlefont in the adjoint representation}
%\medskip
\vskip 1in
\centerline{\bf I.~Jack, D.R.T.~Jones and L.A.~Worthy}
\bigskip
\centerline{\it Department of Mathematical Sciences,  
University of Liverpool, Liverpool L69 3BX, U.K.}
\vskip .3in
We construct a superpotential for the general $\Ncal=\frak12$ 
supersymmetric gauge theory coupled to chiral matter in the adjoint 
representation, and investigate the one-loop renormalisability of the theory.  
\Date{July 2006}}

\newsec{Introduction} $\Ncal=\frak12$ supersymmetric theories (i.e.
theories defined on non-anticommutative superspace) have recently attracted
much attention\ref\ferr{S.~Ferrara and
M.A.~Lledo, JHEP 0005 (2008) 008}\nref\klemm{D.~Klemm, S.~Penati and
L.~Tamassia, Class. Quant. Grav.
20 (2003) 2905}\nref\seib{N.~Seiberg, JHEP
{\bf 0306} (2003) 010}--\ref\araki{T.~Araki, K.~Ito and  A. Ohtsuka,
\plb573 (2003) 209}. Such theories are non-hermitian and
only have half the supersymmetry of the corresponding $\Ncal=1$
theory.
These theories are not power-counting  renormalisable\foot{See
Refs.~\ref\britto{R.~Britto, B.~Feng and S.-J.~Rey,
JHEP 0307 (2003) 067; JHEP 0308 (2003) 001}\ref\terash{S. Terashima
and J-T Yee, JHEP {\bf 0312}
(2003) 053} for other discussions of the ultra-violet properties of these
theories.} but it has been
argued\ref\gris{M.T.~Grisaru, S.~Penati and  A.~Romagnoni, JHEP {\bf
0308} (2003) 003\semi
R.~Britto and B.~Feng, \prl91 (2003) 201601\semi
A.~Romagnoni, JHEP {\bf 0310} (2003) 016}\nref\lunin{O.~Lunin
and S.-J. Rey, JHEP  {\bf 0309}
(2003) 045}\nref\alish{
M.~Alishahiha, A.~Ghodsi and N.~Sadooghi, \npb691
(2004) 111}--\ref\berrey{D.~Berenstein and S.-J.~Rey, \prd68 (2003) 121701}
 that they are in  fact nevertheless
renormalisable, in other words only a finite number of additional
terms need to be added to the lagrangian to absorb divergences to all
orders. In previous work we have confirmed this renormalisability at
the one-loop level. In particular
we have shown that although divergent 
gauge non-invariant terms are generated at the one-loop level, they can be 
removed by  
divergent field redefinitions leading to a renormalisable theory
in which $\Ncal=\frak12$ supersymmetry is preserved at the one-loop level
in both the pure gauge case\ref\jjw{I.~Jack, D.R.T.~Jones
and L.A.~Worthy, \plb611 (2005) 199}\ and in the case of 
chiral matter in the fundamental 
representation\ref\jjwa{I.~Jack, D.R.T.~Jones
and L.A.~Worthy, \prd72 (2005) 065002}. On the other hand, the
authors of Ref.~\ref\penrom{
S.~Penati and A.~Romagnoni, JHEP {\bf 0502} (2005) 064}
obtained the one loop effective action for pure $\Ncal=\frak12$
supersymmetry using a superfield formalism. Although they found
divergent contributions which broke supergauge invariance, their final
result was gauge-invariant without the need for any redefinition. In 
subsequent work\ref\gpr{M.T.~Grisaru, S.~Penati and A.~Romagnoni, 
JHEP {\bf 0602} (2006) 043} it was shown that the $\Ncal=\frak12$
superfield action requires modification to ensure renormalisability, which
is consistent with our findings in the component formulation\jjwa.

It was pointed out in Ref.~\araki\ that an $\Ncal=\frak12$ supersymmetric 
theory can also be constructed with matter in the adjoint representation.
Our purpose here is to repeat the analysis of Ref.~\jjwa\ for the
adjoint case, then proceed to consider the addition of superpotential
terms, which will turn out to be a non-trivial task.
The adjoint action of Ref.~\araki\ was written for the gauge group $U(N)$.
As we noted in Refs.~\jjw, \jjwa, at the quantum level the $U(N)$ gauge 
invariance cannot be retained. In the case of chiral matter in the 
fundamental representation we were obliged to consider a modified theory 
with the gauge group $SU(N)\otimes U(1)$. In the adjoint case with a 
trilinear superpotential, 
it will turn out that the matter fields must also be in a 
representation of $SU(N)\otimes U(1)$. However, for simplicity of exposition we 
shall
start by considering the adjoint case without a superpotential, in other
words adapting the calculations of Ref.~\jjwa\ to the adjoint case.
The classical action without a superpotential may be written
\eqn\Sadj{\eqalign{
S_0=&\int d^4x
\Bigl[-\frak14F^{\mu\nu A}F^A_{\mu\nu}-i\lambdabar^A\sigmabar^{\mu}
(D_{\mu}\lambda)^A+\frak12D^AD^A\cr
&-\frak12iC^{\mu\nu}d^{ABC}e^{ABC}F^A_{\mu\nu}\lambdabar^B\lambdabar^C\cr
&+\frak18g^2|C|^2d^{abe}d^{cde}(\lambdabar^a\lambdabar^b)
(\lambdabar^c\lambdabar^d)
+\frak{1}{4N}\frak{g^4}{g_0^2}|C|^2(\lambdabar^a\lambdabar^a)
(\lambdabar^b\lambdabar^b)\cr
&+
\Fbar F -i\psibar\sigmabar^{\mu}D_{\mu}\psi-D^{\mu}\phibar D_{\mu}\phi\cr
&+g\phibar D^F\phi +
ig\sqrt2(\phibar \lambda^F\psi-\psibar\lambdabar{}^F\phi)\cr
&+d^{abc}gC^{\mu\nu}\left(
\sqrt2D_{\mu}\phibar^a\lambdabar{}^b\sigmabar_{\nu}\psi^c
+i\phibar^a F^b_{\mu\nu}F^c\right)\cr
&+d^{ab0}g_0C^{\mu\nu}\left(
\sqrt2D_{\mu}\phibar^a\lambdabar{}^0\sigmabar_{\nu}\psi^b   
+i\phibar^a F^0_{\mu\nu}F^b\right)\cr
&+d^{000}g_0C^{\mu\nu}\left(
\sqrt2\pa_{\mu}\phibar^0\lambdabar{}^0\sigmabar_{\nu}\psi^0          
+i\phibar^0 F^0_{\mu\nu}F^0\right)\cr      
&+d^{ab0}gC_1^{\mu\nu}\left(\sqrt2\pa_{\mu}\phibar^0\lambdabar{}^a
\sigmabar_{\nu}\psi^b
+i\phibar^0 F^a_{\mu\nu}F^b\right)\cr
&+d^{ab0}gC_2^{\mu\nu}\left(\sqrt2D_{\mu}\phibar^a\lambdabar{}^b
\sigmabar_{\nu}\psi^0
+i\phibar^a F^b_{\mu\nu}F^0\right)\cr
&-\frak{1}{4}g^2|C|^2\phibar \lambdabar^F\lambdabar^FF
\Bigr].\cr}}
Here
\eqn\lamdef{
\lambda^F=\lambda^a\Ftil^a,\quad (\Ftil^A)^{BC}=if^{BAC},}
and we have
\eqn\Dmudef{\eqalign{
D_{\mu}\phi=&\pa_{\mu}\phi+igA^F_{\mu}\phi,\cr
F_{\mu\nu}^A=&\pa_{\mu}A_{\nu}^A-\pa_{\nu}A_{\mu}^A-gf^{ABC}
A_{\mu}^BA_{\nu}^C,\cr}}
with similiar definitions for $D_{\mu}\psi$, $D_{\mu}\lambda$.
If one decomposes $U(N)$ as
$SU(N)\otimes U(1)$ then our convention is that $\phi^a$ (for example) 
are the $SU(N)$
components and $\phi^0$ the $U(1)$ component. (For later convenience 
we also define $g_A$ similarly to encompass both $g_a=g$ and $g_0$.)
Of course then $f^{ABC}=0$
unless all indices are $SU(N)$. 
We note that $d^{ab0}=\sqrt{\frak2N}\delta^{ab}$, $d^{000}=\sqrt{\frak2N}$.
(Useful identities for $U(N)$ are listed in Appendix B.)
We also have
\eqn\etensor{
e^{abc}=g,\quad e^{a0b}=e^{ab0}=e^{000}=g_0,\quad e^{0ab}={g^2\over{g_0}}.}
We have written the $\phibar\lambdabar\lambdabar F$ term as it is 
given starting from the superspace formalism. We note that it has the 
opposite sign from 
that given in Ref.~\araki. This term is $\Ncal=\frak12$ supersymmetric on
its own and so the exact form chosen should not affect the renormalisability
of the theory.
It is easy to show that Eq.~\Sadj\ is invariant under
\eqn\newsusy{
\eqalign{
\delta A^A_{\mu}=&-i\lambdabar^A\sigmabar_{\mu}\epsilon\cr  
\delta \lambda^A_{\alpha}=&i\epsilon_{\alpha}D^A+\left(\sigma^{\mu\nu}\epsilon
\right)_{\alpha}\left[F^A_{\mu\nu}
+\frak12iC_{\mu\nu}e^{ABC}d^{ABC}\lambdabar^B\lambdabar^C\right],\quad
\delta\lambdabar^A_{\alphadot}=0,\cr
\delta D^A=&-\epsilon\sigma^{\mu}D_{\mu}\lambdabar^A,\cr
\delta\phi=&\sqrt2\epsilon\psi,\quad\delta\phibar=0,\cr
\delta\psi^{\alpha}=&\sqrt2\epsilon^{\alpha} F,\quad
\delta\psibar_{\alphadot}=-i\sqrt2(D_{\mu}\phibar)
(\epsilon\sigma^{\mu})_{\alphadot},\cr
\delta F^a=&\delta F^0=0,\cr
\delta \Fbar^a=&-i\sqrt2D_{\mu}\psibar^a\sigmabar^{\mu}\epsilon
-2ig(\phibar\epsilon\lambda^F)^a\cr
&+2gC^{\mu\nu}D_{\mu}(\phibar^b\epsilon\sigma_{\nu}   
\lambdabar^cd^{bca}+\phibar^b\epsilon\sigma_{\nu}
\lambdabar^0d^{b0a})
+2gC_1^{\mu\nu}D_{\mu}(\phibar^0\epsilon\sigma_{\nu}   
\lambdabar^bd^{0ba}),\cr
\delta \Fbar^0=&-i\sqrt2D_{\mu}\psibar^0\sigmabar^{\mu}\epsilon\cr
&+2gC_2^{\mu\nu}D_{\mu}(\phibar^a\epsilon\sigma_{\nu}   
\lambdabar^bd^{ab0})
+2g_0C^{\mu\nu}D_{\mu}(\phibar^0\epsilon\sigma_{\nu}   
\lambdabar^0d^{000}).\cr
}}
In Eq.~\Sadj, $C^{\mu\nu}$ is related to the non-anti-commutativity 
parameter $C^{\alpha\beta}$ by  
\eqn\Cmunu{
C^{\mu\nu}=C^{\alpha\beta}\epsilon_{\beta\gamma}
\sigma^{\mu\nu}_{\alpha}{}^{\gamma},} 
where 
\eqn\sigmunu{\eqalign{
\sigma^{\mu\nu}=&\frak14(\sigma^{\mu}\sigmabar^{\nu}-
\sigma^{\nu}\sigmabar^{\mu}),\cr
\sigmabar^{\mu\nu}=&\frak14(\sigmabar^{\mu}\sigma^{\nu}-
\sigmabar^{\nu}\sigma^{\mu}),\cr }} 
and 
\eqn\Csquar{
|C|^2=C^{\mu\nu}C_{\mu\nu}.} 
Our conventions are in accord with \seib; in particular, 
\eqn\sigid{
\sigma^{\mu}\sigmabar^{\nu}=-\eta^{\mu\nu}+2\sigma^{\mu\nu}.}
Properties of $C$ which follow from
Eq.~\Cmunu\ are  
\eqna\cprop$$\eqalignno{
C^{\alpha\beta}&=\frak12\epsilon^{\alpha\gamma}
\left(\sigma^{\mu\nu}\right)_\gamma{}^{\beta}C_{\mu\nu},
& \cprop a\cr
C^{\mu\nu}\sigma_{\nu\alpha\betadot}&=C_{\alpha}{}^{\gamma}
\sigma^{\mu}{}_{\gamma\betadot},&\cprop b\cr
C^{\mu\nu}\sigmabar_{\nu}^{\alphadot\beta}&=-C^{\beta}{}_{\gamma}
\sigmabar^{\mu\alphadot\gamma}.&\cprop c\cr}$$ 

In Eqs.~\Sadj, $C^{\mu\nu}_{1,2}$ will be identical to $C^{\mu\nu}$ at the
classical level; but we have distinguished them to allow for the
possibility of different renormalisations (in practice an overall numerical
factor) at the quantum level; so that $C^{\mu\nu}_{1,2}$ will obey 
properties analogous to Eqs.~\Cmunu, \Csquar\ and \cprop{}.
It is important to note that this is only 
compatible with $\Ncal=\frak12$ supersymmetry due to the fact that the 
$\pa_{\mu}\phibar^0\lambdabar^a
\sigmabar_{\nu}\psi^b$ term in Eq.~\Sadj\ contains no gauge field; and  
the variation of the gauge field in $D_{\mu}\phibar^a\lambdabar^a
\sigmabar_{\nu}\psi^0$ gives zero. This implies that the variations of the
terms containing either $C^{\mu\nu}_1$ or $C^{\mu\nu}_2$ 
respectively are self-contained. (By contrast, 
the variation of the gauge field in the $D_{\mu}\phibar^a\lambdabar^b
\sigmabar_{\nu}\psi^c$ term is cancelled by the $C^{\mu\nu}$ term in the
variation of the $\lambda$ in the 
$\phibar\lambda\psi$ term, which forces the $C^{\mu\nu}$ in the 6th line of
Eq.~\Sadj\ to be equal to that in the pure gauge terms, and similarly
for that in the 7th line; the terms in the 8th line do not get renormalised
at all.)

We use the standard gauge-fixing term 
\eqn\gafix{
S_{\rm{gf}}={1\over{2\alpha}}\int d^4x (\pa.A)^2} 
with its associated
ghost terms.  The gauge propagators for $SU(N)$ and $U(1)$ are both given by  
\eqn\gprop{
\Delta_{\mu\nu}=-{1\over{p^2}}\left(\eta_{\mu\nu}
+(\alpha-1){p_{\mu}p_{\nu}\over{p^2}}\right)}
(omitting group factors) and the gaugino propagator is  
\eqn\fprop{
\Delta_{\alpha\alphadot}={p_\mu\sigma^{\mu}_{\alpha\alphadot}\over{p^2}},}
where the momentum enters at the end of the propagator with the undotted 
index.  
The one-loop graphs contributing
to the ``standard'' terms in the lagrangian (those without a
$C^{\mu\nu}$) are the same as in the ordinary $\Ncal=1$ case, so 
anomalous dimensions and gauge $\beta$-functions are as for
$\Ncal=1$. Since our gauge-fixing term in Eq.~\gafix\ does not preserve 
supersymmetry, the anomalous dimensions for $A_{\mu}$ and $\lambda$
are
different (and moreover gauge-parameter dependent), as are those for
$\phi$ and $\psi$. However, the 
gauge $\beta$-functions are of course gauge-independent. 
The one-loop one-particle-irreducible (1PI) 
graphs contributing to the new terms (those
containing $C$) are depicted in Figs.~1--6. With the exception of Fig.~6 
(which gives zero contributions in the case of chiral fields in the fundamental
representation) these diagrams are the same as those
considered in Ref.~\jjwa. The divergent contributions from these and other
diagrams considered later are listed in Appendix A. 

\newsec{Renormalisation of the adjoint $SU(N)$ action}
The renormalisation of $\Ncal=\frak12$
supersymmetric gauge theory presents certain subtleties.
The bare action is given by 
\eqn\Sbare{\eqalign{
S_B=&S_{0B}\cr
&+\frak{1}{N}\gamma_1g_0^2|C|^2(\lambdabar^a\lambdabar^a)
(\lambdabar^0\lambdabar^0)\cr
&-gd^{ab0}\gamma_2C_2^{\mu\nu}\left(\sqrt2D_{\mu}\phibar^a\lambdabar^b
\sigmabar_{\nu}\psi^0+\sqrt2\phibar^a\lambdabar^b
\sigmabar_{\nu}\pa_{\mu}\psi^0
+i\phibar^a F^b_{\mu\nu}F^0\right)\cr
&-g_0d^{ab0}\gamma_3C^{\mu\nu}\left(\sqrt2D_{\mu}\phibar^a\lambdabar^0
\sigmabar_{\nu}\psi^b+\sqrt2\phibar^a\lambdabar^0
\sigmabar_{\nu}D_{\mu}\psi^b
+i\phibar^a F^0_{\mu\nu}F^b\right)\cr  
}}
where $S_{0B}$ is obtained by replacing all fields and couplings in $S_0$
(in Eq.~\Sadj) by their bare versions, given below.
The terms involving 
$\gamma_{1-3}$ 
 are separately invariant under $\Ncal=\frak12$ 
supersymmetry. Those with $\gamma_1$, $\gamma_2$
 must be included at this stage to obtain a 
renormalisable lagrangian; those with $\gamma_3$ will be required when we 
introduce a superpotential but could be omitted at present. 

We found in Refs.~\jjw, \jjwa\ that non-linear renormalisations of $\lambda$ 
and $\Fbar$ were required; and in a subsequent
paper\ref\jjwb{I.~Jack, D.R.T.~Jones
and L.A.~Worthy, \prd72 (2005) 107701}\ we pointed out that non-linear 
renormalisations of $F$, $\Fbar$ are required even in ordinary $\Ncal=1$ 
supersymmetric gauge theory when working in the uneliminated formalism.
Note that in the $\Ncal=\frak12$ supersymmetric case, fields and their
conjugates may renormalise differently. 
The renormalisations of the remaining fields and couplings are linear as 
usual and given by
\eqn\bare{\eqalign{ 
\lambdabar^a_B=Z_{\lambda}^{\frak12}\lambdabar^a,
\quad
A^{a}_{\mu B}=Z_A^{\frak12}A^{a}_{\mu}, \quad& D^a_B=Z_D^{\frak12}D^a,
\quad \phi^a_B=Z_{\phi}^{\frak12}\phi^a,\cr
\psi^a_B=Z_{\psi}^{\frak12}\psi^a,\quad
\phibar^a_B=Z_{\phi}^{\frak12}\phibar^a,
\quad \psibar^a_B=&Z_{\psi}^{\frak12}\psibar^a, \quad F_B=Z_FF,\quad
g_B=Z_gg,\cr  
C_B^{\mu\nu}=Z_CC^{\mu\nu}, \quad |C|_B^2=Z_{|C|^2}|C|^2,\quad&
C_{1,2B}^{\mu\nu}=Z_{C_{1,2}}C_{1,2}^{\mu\nu},\quad \gamma_{1-3B}=
Z_{1-3}.
\cr}}
The corresponding $U(1)$ gauge multiplet fields 
$\lambdabar^0$ etc are unrenormalised 
(as are the $U(1)$ chiral fields $\phi^0$ etc in the case with no 
superpotential); so is $g_0$. The auxiliary field $F$ is also unrenormalised,
i.e. $Z_F=1$
(though again this will no longer be the case when we later introduce a 
superpotential). In Eq.~\bare, $Z_{1-3}$ are divergent
contributions, in other words we have set the renormalised couplings
$\gamma_{1-3}$ to zero for simplicity.   
The other renormalisation constants start with 
tree-level values of 1. As we mentioned before,
the renormalisation constants for the fields
and for the gauge coupling $g$ are the same as in the ordinary $\Ncal=1$
supersymmetric theory (for a gauge theory coupled to an adjoint
chiral field) and are therefore given up to one loop 
by\ref\timj{
D.~Gross and F.~Wilcek, \prd8 (1973) 3633\semi
D.R.T.~Jones, \npb87 (1975) 127}: 
\eqn\Zgg{\eqalign{
Z_{\lambda}=&1-g^2NL(2\alpha +2),\cr
Z_A=&1+g^2NL(1-\alpha)\cr
Z_D=&1-2NLg^2,\cr
Z_g=&1-2g^2NL,\cr
Z_{\phi}=&1+2g^2(1-\alpha)LN,\cr
Z_{\psi}=&1-2g^2(1+\alpha)LN,\cr}}
where (using dimensional regularisation with $d=4-\epsilon$)
$L={1\over{16\pi^2\epsilon}}$.
The renormalisation of $\lambda^A$ is given by 
\eqn\lchange{\eqalign{
\lambda_B^a=&Z_{\lambda}^{\frak12}\lambda^a
-\frak12NLg^3C^{\mu\nu}d^{abc}
\sigma_{\mu}\lambdabar^cA_{\nu}^b
-NLg^2g_0C^{\mu\nu}d^{ab0}         
\sigma_{\mu}\lambdabar^0A_{\nu}^b\cr
&+i\sqrt2\rho_4NLg^3d^{abc}(C\psi)^b\phibar^c
+i\sqrt2\rho_5NLg^3d^{ab0}(C\psi)^0\phibar^b,\cr
\lambda_B^0=&\lambda_0i\sqrt2\rho_6NLg^2g_0d^{0ab}
(C\psi)^a\phibar^b,\cr 
}}
where $(C\psi)^{\alpha}=C^{\alpha}{}_{\beta}\psi^{\beta}$. The replacement
of $\lambda$ by $\lambda_B$ produces a change in the action given (to
first order) by 
\eqn\Schlamb{\eqalign{
S_0(\lambda_B)-S_0(\lambda)=&NLg^2\int d^4x\Bigl\{
\rho_4g\bigl[igd^{abe}f^{cde}\phibar^a\phibar^b\psi^c(C\psi^d)\cr
&+\sqrt2C^{\mu\nu}d^{abc}\phi^a\lambdabar^b\sigmabar_{\nu}D_{\mu}\psi^c
+\sqrt2C^{\mu\nu}d^{abc}D_{\mu}\phibar^a\lambdabar^b\sigmabar_{\nu}
\psi^c\bigr]\cr
&+\rho_5\sqrt2gC^{\mu\nu}d^{ab0}(\phibar^a\lambdabar^b\sigmabar_{\nu}
\pa_{\mu}\psi^0
+D_{\mu}\phibar^a\lambdabar^b\sigmabar_{\nu}\psi^0)\cr
&+\rho_6\sqrt2g_0C^{\mu\nu}d^{0ab}(\phibar^a\lambdabar^0\sigmabar_{\nu}
D_{\mu}\psi^b
+D_{\mu}\phibar^a\lambdabar^0\sigmabar_{\nu}\psi^b)+\ldots\Bigr\},\cr}}
where the ellipsis indicates the terms not involving $\rho_{4-6}$ 
(which were given previously in Ref.~\jjwa). 
The value of $\rho_4$ will be chosen
so as to cancel the divergent contributions from Fig.~6; $\rho_{5,6}$
will be specified later when we renormalise the theory with a 
superpotential. 

We now find that to render finite the contributions linear in $F$ we require
\eqn\fbarredef{\eqalign{
\Fbar^a_B=&Z_F\Fbar^a+
iC^{\mu\nu}Lg^2\Bigl\{gN\Bigl[(5+2\alpha)
\pa_{\mu}A_{\nu}^b-\frak14(11+4\alpha)
gf^{bde}A_{\mu}^dA_{\nu}^e\Bigr]\phibar^c d^{abc}\cr
&+\sqrt{2N}g\Bigl[2\left((4+\alpha)-z_{C_1}\right)\pa_{\mu}A_{\nu}^a
-\left(\frak12(9+2\alpha)-z_{C_1}\right)
gf^{abc}A_{\mu}^bA_{\nu}^c\Bigr]\phibar^0\cr
&+2\sqrt{2N}g_0\left(-(1-\alpha)+z_3\right)
\pa_{\mu}A_{\nu}^0\phibar^a\Bigr\}\cr
&+\frak{1}{8}Lg^4|C^2|\Bigl[2(1-\alpha)Nf^{ace}f^{bde}
-11Nd^{abe}d^{cde}+4(\delta^{ab}\delta^{cd}+\delta^{ac}\delta^{bd})\Bigr]
\phibar^b\lambdabar^c\lambdabar^d\cr
&-Lg^3|C|^2\Bigl\{d^{abc}\sqrt{2N}\Bigl[g\phibar^0\lambdabar^b\lambdabar^c
+3g_0\phibar^b\lambdabar^c\lambdabar^0\Bigr]
+4g_0\phibar^0\lambdabar^0\lambdabar^a\Bigr\},\cr
\Fbar^0_B=&Z_F\Fbar^0
+i\sqrt{2N}L\left(2+z_2-z_{C_2}\right)C^{\mu\nu}F^a_{\mu\nu}\phibar^a\cr
&-2d^{abc}g^4L|C|^2\sqrt{2N}\phibar^a\lambdabar^b\lambdabar^c
-8g^3g_0L|C|^2\phibar^a\lambdabar^a\lambdabar^0.\cr}}
Writing $Z_C^{(n)}$ for the $n$-loop
contribution to $Z_C$ we set
\eqn\zformb{
Z_C^{(1)}=z_CNLg^2}
with similar definitions for $Z_{|C|^2}$, $Z_{C_{1,2}}$, $Z_{1-3}$.
We now find that with
\eqn\zforms{
z_C=z_{|C|^2}=0,\quad z_{C_1}=-z_{C_2}=2,\quad
z_1=-3,\quad \rho_4=1,\quad \rho_5=z_2-z_{C_2}, \quad \rho_6=z_3,}
the one-loop effective action is finite, for arbitrary $z_2$, $z_3$.. 

\newsec{The superpotential}
We now consider the problem of adding superpotential terms to the 
lagrangian Eq.~\Sadj. The following potential terms are $\Ncal=\frak12$ 
invariant at the classical level:  
\eqn\lagmass{\eqalign{
S_{\rm{int}}=&\int d^4x\tr\Bigl\{y\bigl[
\phi^2 F-\psi^2\phi\cr
&+\phibar^2\Fbar-\psibar^2\phibar+\frak43ig
C^{\mu\nu}\phibar^3\Fhat_{\mu\nu}+\frak23C^{\mu\nu}D_{\mu}\phibar D_{\nu}\phibar
\phibar\bigr]\cr
&+m\bigl[\phi F -\frak12\psi\psi+\phibar\Fbar-\frak12\psibar\psibar
+iC^{\mu\nu}\phibar \Fhat_{\mu\nu}\phibar
-\frak18g^3|C|^2\phibar \phibar\lambdabar^F
\lambdabar^F\bigr]\Bigr\}.\cr}}
Here in the interests of conciseness we have written the superpotential
in index-free form, so that 
\eqn\hatdef{
\phi=\phi^A R^A, \quad 
\psi=\psi^AR^A,\quad \Ahat_{\mu}=gA_{\mu}^aR^a+g_0A_{\mu}^0R^0;}
it then follows that $\Fhat_{\mu\nu}=g_AF_{\mu\nu}^AR^A$, with 
$F_{\mu\nu}$ defined as in Eq.~\Dmudef. 
The group matrices are normalised so that
$\Tr[R^AR^B]=\frak12\delta^{AB}$; in particular, $R^0=\sqrt{\frak{1}{2N}}1$.
It is easy to check that $S_{\rm{int}}$ is $\Ncal=\frak12$ invariant.
Except for the last mass term, this superpotential is most readily  
derived directly from the superspace formalism.
Denoting an adjoint chiral superfield as 
$\Phi_A$, 
we have that under a gauge transformation 
$$
\Phi_A \to \Omega * \Phi_A *  \Omega^{-1}, \quad 
\ov \Phi_A \to \ov \Omega * \ov \Phi_A *  \ov \Omega^{-1}, 
$$
so that the gauge interactions are written in superfield form as 
$$
\int d^4\th \,\tr\left[\ov \Phi_A * e^V * \Phi_A * e^{-V}\right]  .
$$
The following superpotential terms are manifestly also invariant:
\eqn\super{\eqalign{
&\int d^2 \theta 
\,\tr\left[\frak12m \Phi_A * \Phi_A + \frak13y\Phi_A * \Phi_A * \Phi_A  
\right]  
\quad \cr
+&  
\int d^2 \bth \,\tr\left[\frak13m \ov \Phi_A * \ov \Phi_A 
+ \frak13y\ov \Phi_A * \ov \Phi_A * \ov \Phi_A  \right].
}}

Expanded in component fields we have 
\eqna\comps$$\eqalignno{
\Phi_A (y, \th) &= \phi (y) + \sqrt{2} \th \psi (y) + \th\th
F (y) & \comps a\cr
\ov \Phi_A (\ov y, \bth) &= \ov \phi (\ov y) + \sqrt{2} \bth \ov \psi (\ov y)\cr
&+ \bth\bth \Big( \ov F (\ov y) + i gC^{\mu\nu} \partial_{\mu}
\{\ov \phi, A_{\nu}\}(\ov y) 
- \frak{g^2}{2} C^{\mu\nu} \left[ A_{\mu},\{ A_{\nu},  
\ov \phi\}\right](\ov y) \Big), & 
\comps b\cr
}$$
where $\ov y^{\mu} = y^{\mu} - 2 i \th \sigma^{\mu} \bth$. Note the 
modification of the $\bth\bth$-term\araki.

If we substitute Eq.~\comps{}\ in Eq.~\super\ we obtain Eq.~\lagmass\
except for the 
last term. (This can also
be expressed in superfields but in a more unwieldy form). 
The coefficient 
of this final term is arbitrary since it is separately 
$\Ncal=\frak12$ invariant; the reason for our particular choice will be
explained later (after Eq.~(A.18) in Appendix A).
A similar set of mass terms is admissible in the case of the 
fundamental representation, with mass terms coupling the fundamental and
anti-fundamental representation fields\ref\jjwc{
I.~Jack, D.R.T.~Jones and L.A.~Worthy, ``One-loop renormalisation of massive
$\Ncal=\frak12$ supersymmetric gauge theory'' in
preparation}. However, no trilinear term is possible
in the $\Ncal=\frak12$ case for the fundamental representation.
If we have both adjoint and fundamental (antifundamental) representations 
$\Phi (\wt \Phi)$ we can construct \ntwo-type invariants, of the form 
\eqn\supertwo{
y \left[\int d^2 \theta \,\wt \Phi * \Phi_A * \Phi \quad + \quad  
\int d^2 \bth \,\ov \Phi * \ov \Phi_A * \ov {\wt \Phi} \right].
}

At the classical level $\phi$ may be 
considered as forming a representation of $U(N)$. However, just as we saw in
Ref.~\jjwa\ for the gauge group, the $U(N)$ structure is not preserved at the
quantum level. The $\phi^a$ renormalise differently from the $\phi^0$ and this 
means that, for instance, there must be a different mass parameter ($m$, say)
for the $\phi^a F^a$, $\psi^a\psi^a$ terms than for the $\phi^0 F^0$, 
$\psi^0\psi^0$ terms ($m_0$, say). In the case of the mass terms this does not 
present serious difficulty since we can separate the mass terms in 
Eq.~\lagmass\ into separately $\Ncal=\frak12$ invariant sets of terms
involving either $m$ or $m_0$. However, in the case of the trilinear 
superpotential terms, we need to invoke three separate couplings, one
($y$, say) for $\phi^a\phi^b F^c$ terms, one ($y_1$, say)
for $\phi^a\phi^b F^0$, $\phi^a\phi^0 F^b$ etc and one ($y_2$, say)
for $\phi^0\phi^0 F^0$. In the $\Ncal=1$ case the theory would, of course,
be renormalisable, with each of $y$, $y_{1,2}$ 
renormalising differently. By contrast, in the $\Ncal=\frak12$ case many of the 
$\phibar^3 A_{\mu}$ terms are linked by $\Ncal=\frak12$ transformations 
to more than one of these groups of terms and so cannot be assigned a unique
coupling out of $y$, $y_{1,2}$. So in the presence of trilinear 
superpotential terms, the $\Ncal=\frak12$ invariance cannot be maintained at 
the quantum level. It is this linking of different groups of terms, 
specifically those corresponding purely to $SU(N)$ with those containing
$U(1)$ fields, which implies that we cannot have an $\Ncal=\frak12$ theory
with a superpotential if the chiral fields belong to $SU(N)$ alone. 
\vfill
\eject
\newsec{The renormalised action with superpotential}
As we explained in the previous section, many of the individual terms
with couplings $m$ or $y$ 
in Eq.~\lagmass\ will renormalise differently and hence need to
be assigned their own separate couplings. 
For renormalisability, Eq.~\lagmass\ needs to be replaced by
\eqn\lagmassren{\eqalign{
S_{\rm{int}}=&\int d^4x\Bigl\{\frak14y d^{abc}(\phi^a\phi^b F^c
-\psi^a\psi^b\phi^c)\cr
&+\frak14y_1d^{ab0}(\phi^a\phi^b F^0
+2\phi^a\phi^0F^b-\psi^a\psi^b\phi^0-2\psi^a\psi^0\phi^b)\cr
&+\frak14y_2d^{000}(\phi^0\phi^0F^0-\psi^0\psi^0\phi^0)
\cr
&+\frak14y d^{abc}(\phibar^a\phibar^b \Fbar^c
-\psibar^a\psibar^b\phibar^c)\cr          
&+\frak14y_1d^{ab0}(\phibar^a\phibar^b \Fbar^0+2\phibar^a\phibar^0\Fbar^b
-\psibar^a\psibar^b\phibar^0 
-2\psibar^a\psibar^0\phibar^b)\cr
&+\frak14y_2d^{000}(\phibar^0\phibar^0\Fbar^0
-\psibar^0\psibar^0\phibar^0)
\cr
&+igC^{\mu\nu}F^d_{\mu\nu}\Bigl(
\frak16y d^{abe}d^{cde}\phibar^a\phibar^b\phibar^c
+\frak13y_3\frak1N\delta^{ab}\delta^{cd}\phibar^a\phibar^b\phibar^c\cr
&+\frak12y_4\sqrt{\frak2N}d^{abd}\phibar^0\phibar^a\phibar^b
+y_5\frak1N\phibar^0\phibar^0\phibar^d\Bigr)\cr
&+ig_0C^{\mu\nu}F_{\mu\nu}^0\Bigl(\frak16y\sqrt{\frak2N}
d^{abc}\phibar^a\phibar^b\phibar^c
+y_1\frak1N\phibar^a\phibar^a\phibar^0
+\frak13y_2\frak1N\phibar^0\phibar^0\phibar^0\Bigr)\cr
&+\frak16iy C^{\mu\nu}f^{abc}(D_{\mu}\phibar)^a (D_{\nu}\phibar)^b
\phibar^c\cr
&+m\Bigl[\phi^a F^a -\frak12\psi^a\psi^a+\phibar^a\Fbar^a
-\frak12\psibar^a\psibar^a\cr
&+\frak12igC^{\mu\nu}d^{abc}F^c_{\mu\nu}\phibar^a \phibar^b
+\frak12ig
C_1^{\mu\nu}d^{0ab}F^a_{\mu\nu}\phibar^0 \phibar^b
+\frak12ig
C^{\mu\nu}d^{0ab}F^0_{\mu\nu}\phibar^a\phibar^b
\Bigr]\cr
&+m_0\Bigl[\phi^0 F^0 -\frak12\psi^0\psi^0+\phibar^0\Fbar^0
-\frak12\psibar^0\psibar^0\cr
&+\frak12igC_2^{\mu\nu}d^{0ab}F^a_{\mu\nu}\phibar^0 \phibar^b
+\frak12ig
C^{\mu\nu}d^{000}F^0_{\mu\nu}\phibar^0 \phibar^0\Bigr]\cr
&+|C|^2\bigl[-\frak18
\mu_1f^{ace}f^{bde}+\mu_2\frak2N\delta^{ab}\delta^{cd}\bigr]
g^2\phibar^a\phibar^b\lambdabar^c\lambdabar^d\cr
&+gd^{abc}\sqrt{2N}|C|^2\phibar^a\lambdabar^b\left(\mu_3g_0\phibar^c\lambdabar^0
+g\mu_4\phibar^0\lambdabar^c\right)
+\frak2N\mu_5gg_0|C|^2\phibar^a\phibar^0\lambdabar^a\lambdabar^0
\Bigr\}.\cr}}
Each of the coefficients $m$, $y$, etc above will renormalise 
separately. However, for simplicity when we quote the results for
Feynman diagrams, we will use the values of the coefficients as implied by
Eq.~\lagmass, i.e. $y_{1-5}=y$, $m_0=\mu_1=m$, $\mu_{2-5}=0$,
so that these are effectively        
the renormalised values of these couplings.
Note that the $g_0C^{\mu\nu}F_{\mu\nu}^0\sqrt{\frak2N}
d^{abc}\phibar^a\phibar^b\phibar^c$ and 
$\frak1Ng_0C^{\mu\nu}F_{\mu\nu}^0\phibar^a\phibar^a\phibar^0$
terms only mix with the $d^{abc}\phibar^a\phibar^b\Fbar^c$ or 
$d^{ab0}\phibar^a\phibar^b\Fbar^0$ fields respectively and hence
can be assigned the coupling $y$ or $y_1$ respectively.

The renormalisation constants $Z_{\phi,\psi}$, $Z_F$
now acquire $y$-dependent contributions, so we have 
\eqn\newZ{
\eqalign{
Z_{\phi}=&1+\left[-\frak14y^2+2g^2(1-\alpha)\right]LN,\cr
Z_{\psi}=&1+\left[-\frak14y^2-2g^2(1+\alpha)\right]LN,\cr
Z_{\phi^0}=&Z_{\psi^0}=1-\frak14y^2LN,\cr
Z_F=&1-\frak14y^2LN.\cr}}
Here we write $\phi^0_B=Z_{\phi^0}^{\frak12}\phi^0$, etc, since the
$U(1)$ chiral fields are now renormalised.
Now for the bare action we also need to replace $m_B=Z_m m$, 
$y_B=Z_{y} y$ etc in addition to the replacements given
earlier. These renormalisation constants are given according to the
non-renormalisation theorem by
\eqn\zlam{\eqalign{
Z_m=&Z_{\Phi}^{-1},\cr
Z_{m_0}=&Z_{\Phi_0}^{-1},\cr
Z_{y}=&Z_{\Phi}^{-\frak32},\cr
Z_{y_1}=&Z_{\Phi}^{-1}Z_{\Phi_0}^{-\frak12},\cr
Z_{y_2}=&Z_{\Phi_0}^{-\frak32},\cr}}
where $Z_{\Phi}$ , $Z_{\Phi_0}$ are the
renormalisation constants for the chiral superfield $\Phi$ given by
\eqn\zPhi{
\eqalign{
Z_{\Phi}=&1+\left[-\frak14y^2+4g^2\right]LN,\cr
Z_{\Phi^0}=&1-\frak14y^2LN.\cr}}
 
The redefinitions of $F$ and $\Fbar$ found in Ref.~\jjw\ need to be
modified in the presence of mass terms and the $U(1)$ gauge group. 
This is easily done following the
arguments of Ref.~\jjwb; there are no one-loop diagrams giving divergent 
contributions to $m\phi F$ or $m\phibar\Fbar$ although there are counterterm
contributions from $m_B\phi_B F$, $m_B\phibar_B\Fbar$. 
At one loop we have 
\eqn\newF{\eqalign{
\Fbar^{a\prime}_B=&\Fbar^{a}_B+(\alpha+3)g^2NL\left(m\phibar^a
+\frak14y d^{abc}\phibar^b\phibar^c\right) 
+\frak12(\alpha+3)y g^2NLd^{ab0}\phibar^b\phibar^0,\cr
\Fbar_B^{0\prime}=&\Fbar^0_B,\cr
F^{a\prime}_B=&Z_FF^a+(\alpha+3)g^2NL\left(m\phi^a
 +\frak14y d^{abc}\phi^b\phi^c\right) 
+\frak12(\alpha+3)y g^2NLd^{ab0}\phi^b\phi^0,\cr
F^{0\prime}_B=&Z_FF^0.\cr}}
Here $\Fbar^{a}_B$ etc are as given in Eq.~\fbarredef, though of course using 
the non-zero $Z_F$ given in Eq.~\newZ. 
The new $C$-dependent diagrams in the presence of a superpotential 
are depicted in Figs.~7--11, and their divergent contributions in the
corresponding Tables. We omit diagrams giving contributions of the form 
$A_{\mu}A_{\nu}\phibar^3$ which complete the $F_{\mu\nu}$ in 
$F_{\mu\nu}\phibar^3$
contributions; we already have ample evidence that gauge invariance, even
when apparently violated, can be restored by making divergent field 
redefinitions. We also omit diagrams of the form 
$\phibar^3\lambdabar^2$; these are separately $\Ncal=\frak12$ invariant and
are not going to give any more information about the preservation
of $\Ncal=\frak12$ supersymmetry.
 
We now choose the renormalisation constants at our disposal to
ensure finiteness. In order to ensure renormalisability of the action in 
Eq.~\lagmassren, we find we now need to impose specific values for the hitherto
arbitrary coefficients $z_2$, $z_3$, namely
\eqn\zformb{
z_2=-4,\quad z_3=4.}
We find moreover
\eqn\moreZ{\eqalign{
Z_{y_3}=&1-6LNg^2,\cr
Z_{y_4}=&1-4LNg^2,\cr
Z_{y_5}=&1-2LNg^2,\cr
Z_{\mu_1}=&1+\frak{32}{N}Lg^2\left(1-\frak{g^2}{g_0^2}\right),\cr
Z_{\mu_2}=&-\frak{g^4}{g_0^2}L,\cr
Z_{\mu_3}=&0,\cr
Z_{\mu_4}=&2LNg^2,\cr
Z_{\mu_5}=&4LNg^2.\cr
}}
\vfill
\eject
\newsec{The eliminated formalism}
It is instructive and also provides a useful check to perform the calculation 
in the eliminated formalism. In the eliminated case Eq.~\lagmassren\ is replaced
by
\eqn\lagmassa{\eqalign{
\tilde S_{\rm{mass}}=&\int d^4x\Bigl\{
-\frak14y d^{abc}\psi^a\psi^b\phi^c
-\frak14y_1d^{ab0}(\psi^a\psi^b\phi^0+2\psi^a\psi^0\phi^b)
-\frak14y_2d^{000}\psi^0\psi^0\phi^0\cr
&-\frak14y d^{abc}\psibar^a\psibar^b\phibar^c       
-\frak14y_1d^{ab0}(\psibar^a\psibar^b\phibar^0
+2\psibar^a\psibar^0\phibar^b)
-\frak14y_2d^{000}\psibar^0\psibar^0\phibar^0                  
\cr
&+m^2\phibar^a\phi^a
-\frak12m\psi^a\psi^a-\frak12m\psibar^a\psibar^a
+m_0^2\phibar^0\phi^0      
-\frak12m_0\psi^0\psi^0-\frak12m_0\psibar^0\psibar^0
\cr
&-\left(y d^{eab}\phi^a\phi^b+2y_1 d^{ea0}\phi^a\phi^0\right)
\left(y d^{eab}\phibar^a\phibar^b+2y_1 
d^{ea0}\phibar^a\phibar^0\right)\cr
&-\left(y_1 d^{0ab}\phi^a\phi^b+y_2 d^{000}\phi^0\phi^0\right)
\left(y_1 d^{0ab}\phibar^a\phibar^b+
y_2 d^{000}\phibar^0\phibar^0\right)\cr
&+\frak16iy C^{\mu\nu}f^{abc}(D_{\mu}\phibar)^a (D_{\nu}\phibar)^b
\phibar^c\cr
&+\frak16igF^d_{\mu\nu}\Bigl(
-\frak12y C^{\mu\nu}d^{abe}d^{cde}\phibar^a\phibar^b\phibar^c
+[2y_3C^{\mu\nu}-3y_1(1-Z_2)C_2^{\mu\nu}]\frak1N
\delta^{ab}\delta^{cd}\phibar^a\phibar^b\phibar^c\cr
&+3[(y_4-y_1)C^{\mu\nu}-\frak12y C_1^{\mu\nu}]
\sqrt{\frak2N}d^{abd}\phibar^0\phibar^a\phibar^b\cr
&+6[y_5C^{\mu\nu}-y_1C_1^{\mu\nu}
-\frak12y_2(1-Z_2)C_2^{\mu\nu}]\frak1N\phibar^0\phibar^0\phibar^d\Bigr)\cr
&-\frak{1}{12}ig_0C^{\mu\nu}F_{\mu\nu}^0\Bigl(y
(1-3Z_3)\sqrt{\frak2N}
d^{abc}\phibar^a\phibar^b\phibar^c
+6y_1(1-2Z_3)
\frak1N\phibar^a\phibar^a\phibar^0\cr
&+2y_2\frak1N\phibar^0\phibar^0\phibar^0\Bigr)\cr
&-\frak12i\Bigl\{gmd^{abc}C^{\mu\nu}F^c_{\mu\nu}\phibar^a \phibar^b
+g_0m_0(1-2Z_3)C^{\mu\nu}d^{ab0}F^0_{\mu\nu}\phibar^a \phibar^b\cr
&+g\left[mC_1^{\mu\nu}
+m_0(1-2Z_2)C_2^{\mu\nu}\right]
d^{ab0}F^b_{\mu\nu}\phibar^a \phibar^0\Bigr\}\cr
&+g^2|C|^2\bigl[-\frak18(\mu_1-2m)f^{ace}f^{bde}
+\mu_2\frak2N\delta^{ab}\delta^{cd}\bigr]
\phibar^a\phibar^b\lambdabar^c\lambdabar^d\cr
&+gd^{abc}\sqrt{2N}|C|^2\phibar^a\lambdabar^b\left(\mu_3g_0\phibar^c\lambdabar^0
+g\mu_4\phibar^0\lambdabar^c\right)
+\frak2N\mu_5gg_0|C|^2\phibar^a\phibar^0\lambdabar^a\lambdabar^0\Bigr\}\cr
}}
while we simply strike out the terms involving $F$, $\Fbar$ in Eq.~\Sadj.
Once again note that in quoting diagrammatic results we set
$y_{1-5}=y$, $m_0=\mu_1=m$, $\mu_{2-5}=0$, so that these are effectively
the renormalised values of these couplings. In Table~7, the contributions from 
Figs.~7(f-k) are now absent while those from
Figs.~7(l-r) change sign. Similarly, in Table~8, the contributions from 
Figs.~8(e-p) are now absent while those from Figs.~8(q-dd) change sign.
In Table~9, the contributions from Figs.~9(f-n) are now absent while
those from Figs.~9(o-z) change sign. In Table~10, the contribution from 
Fig.~10(d) is now absent. In Table~11, the contributions from 
Figs.~11(j-o) are now absent while those from Figs.~11(p-v) which contain 
two factors of $d^{abc}$ acquire an additional factor of 
$\left(-\frak12\right)$.
The results from Figs.~7--11 now add to
\eqn\sumsixelim{\eqalign{
\Gamma_{7\rm{1PI elim}}^{(1)\rm{pole}}=
&iLg^2C^{\mu\nu}m\Bigl[-\frak12(7+5\alpha)Ngd^{abc}\pa_{\mu}A_{\nu}^a\phibar^b
\phibar^c\cr
&+3(1-\alpha)g\sqrt{2N}\pa_{\mu}A_{\nu}^a\phibar^a\phibar^0
-2(5+\alpha)g_0\sqrt{2N}\pa_{\mu}A_{\nu}^0\phibar^a\phibar^a\Bigr],\cr
\Gamma_{8\rm{1PI elim}}^{(1)\rm{pole}}=&
iLg^4C^{\mu\nu}mf^{abe}A_{\mu}^aA_{\nu}^b\Bigl[
\frak12(5+3\alpha)Nd^{cde}\phibar^c
\phibar^d+2\alpha\sqrt{2N}\phibar^e\phibar^0\Bigr],\cr
\Gamma_{9\rm{1PIelim}}^{(1)\rm{pole}}=&
|C|^2mL\Bigl\{\bigl[
2\frak{g^2}{g_0^2}\delta^{ab}\delta^{cd}
+\left\{\frak12N(3+\alpha)+\frak4N\left(1-\frak{g^2}{g_0^2}\right)\right\}
f^{ace}f^{bde}\bigr]
g^4\phibar^a\phibar^b\lambdabar^c\lambdabar^d\cr
&-2g^4d^{abc}\sqrt{2N}\phibar^a\lambdabar^b\phibar^0\lambdabar^c
-8g^3g_0\phibar^a\phibar^0\lambdabar^a\lambdabar^0\Bigr\},\cr
\Gamma_{10\rm{1PI elim}}^{(1)\rm{pole}}=&
\Gamma_{10\rm{1PI}}^{(1)\rm{pole}},\cr
\Gamma_{11\rm{1PI elim}}^{(1)\rm{pole}}=&iC^{\mu\nu}\lambda g^2L\Bigl(
-\frak12g\left(3+\frak73\alpha\right)Nf^{abe}f^{cde}\pa_{\mu}\phibar^a
\phibar^b\phibar^cA_{\nu}^d\cr
&+\left[-\left(\frak34+\frak{7}{12}\alpha\right)d^{abe}d^{cde}
+\left(\frak52-\frak76\alpha\right)\delta^{ab}\delta^{cd}\right]
g\phibar^a\phibar^b\phibar^c\pa_{\mu}A_{\nu}^d\cr
&-\frak14(7+5\alpha)g\sqrt{2N}d^{abc}\phibar^0\phibar^a\phibar^b
\pa_{\mu}A_{\nu}^c+\frak32(1-\alpha)g\phibar^0\phibar^0\phibar^a 
\pa_{\mu}A_{\nu}^a\cr
&-\frak12(5+\alpha)g_0\sqrt{2N}d^{abc}\phibar^a\phibar^b\phibar^c\pa_{\mu}A_{\nu}^0
-2(5+\alpha)g_0\phibar^a\phibar^a\phibar^0\pa_{\mu}A_{\nu}^0\Bigr),\cr
}}
respectively. The results in Eq.~\moreZ\ are unchanged, which is a very good
check on the calculation.

\newsec{Conclusions}
We have repeated our earlier one-loop analysis of $\Ncal=\frak12$ supersymmetry
for the case of chiral matter in the adjoint representation.
We have constructed an $\Ncal=\frak12$ invariant set of mass terms and an
$\Ncal=\frak12$ invariant set of trilinear terms 
for this case. The $\Ncal=\frak12$ invariance of the trilinear terms
requires the chiral matter be in the adjoint representation of 
$U(N)$ rather than $SU(N)$ at the classical level. However, once we consider
quantum corrections, the $U(1)$ chiral fields will renormalise differently
from the $SU(N)$ fields and so at the quantum level we are obliged to
consider $SU(N)\otimes U(1)$ rather than $U(N)$.  On the other hand, the
$\Ncal=\frak12$ transformations mix superpotential terms with different
kinds of field ($SU(N)$ or $U(1)$) and so it is clear that the 
$\Ncal=\frak12$ invariance of the trilinear terms cannot be preserved 
at the quantum level. We have shown
that the $\Ncal=\frak12$ supersymmetry of the mass terms is preserved under 
renormalisation at the one-loop level, and also that certain groups of
trilinear terms for which $\Ncal=\frak12$ supersymmetry does not mix the 
different gauge groups remain $\Ncal=\frak12$ supersymmetric at one loop.
However the renormalisability is assured by making a
particular choice of the parameters $\gamma_2$, $\gamma_3$
(in Eq.~\Sbare), as determined by Eq.~\zformb. 
This also implies (through Eq.~\zforms) a particular choice of renormalisation
for the gaugino $\lambda$, parametrised by $\rho_5$  (in Eq.~\lchange). 
The necessity for these choices seems somewhat counterintuitive as these
renormalisations are all present in the theory without 
superpotential and yet there
appeared to be nothing in the theory without       
superpotential to enforce these choices.
It would be reassuring if some independent confirmation could be found for
these particular values. Presumably the necessity for the non-linear
renormalisations we are compelled to make lies in our use of a
non-supersymmetric gauge (the obvious choice when working in components, of
course). So the answer to this puzzle might lie in a close scrutiny of the
gauge-invariance Ward identities. Of course a calculation in superspace
would also be illuminating. It is always tempting to investigate whether
the behaviour at one loop persists to higher orders but the proliferation
of diagrams in this case would almost certainly be prohibitive.

\centerline{{\bf Acknowledgements}}\nobreak

LAW was supported by PPARC through a Graduate Studentship. 

\appendix{A}{Results for one-loop diagrams}
In this Appendix we list the divergent contributions from the various one-loop
diagrams. 

The contributions from the graphs shown in Fig.~1 are of the form
\eqn\formone{
\sqrt2Ng^2g_BLC^{\mu\nu}d^{ABC}\left(\pa_{\mu}\phibar^A X_1^{ABC}\lambdabar^B 
\sigmabar_{\nu}\psi^C
+\phibar^A Y_1^{ABC}\lambdabar^B \sigmabar_{\nu}\pa_{\mu}\psi^C\right)  }
where 
$X_1^{ABC}$ and $Y_1^{ABC}$ consist of a number $X_1$, $Y_1$ multiplying a 
tensor structure formed of a product of terms like $c^A$ or $d^A$,
where $c^A=1-\delta^{A0}$, $d^A=1+\delta^{A0}$.
 The 
$X_1$, $Y_1$ and the tensor structures are given separately in Table~1. 
(The contributions from Figs.~2--4, 7, 8 also involve tensors 
$X^{ABC\ldots}_i$, $Y_i^{ABC\ldots}$ etc (for Fig.~$i$)
which can be decomposed similarly and will be similarly presented.)
\bigskip
\vbox{
\begintable
Fig.| $X_1$| $Y_1$|Tensor\cr
1a|$\frak32$|$-\alpha $|$c^Ac^Bd^C$\cr
1b|$\alpha $|$\alpha $|$c^Ac^Bd^C$\cr
1c|$\alpha $|$0$|$d^Ac^Bc^C$\cr
1d|$1$|$-1$|$c^Ac^Bd^C$\cr
1e|$1$|$0$|$d^Ac^Bc^C$\cr
1f|$-\frak12 (1-2\alpha)$|$0$|$c^Ad^Bc^C$\cr
1g|$1$|$0$|$c^Ad^Bc^C$\cr
1h|$1$|$-1$|$c^Ad^Bc^C$\cr
1i|$0$|$1$|$c^Ad^Bc^C$\cr
1j|$-3$|$0$|$c^A$
\endtable}
\centerline{{\it Table~1:\/} Contributions from Fig.~1}
\bigskip

The sum of the contributions from Table~1 can be written in the form
\eqn\sumone{\eqalign{  
\Gamma_{1\rm{1PI}}^{(1)\rm{pole}}=&
Ng^2\sqrt2LC^{\mu\nu}\Bigl[(2+3\alpha)gd^{abc}\pa_{\mu}\phibar^a 
\lambdabar^b\sigmabar_{\nu}\psi^c
-gd^{abc}\phibar^a \lambdabar^b \sigmabar_{\nu}\pa_{\mu}\psi^c  \cr
& +2(1+\alpha)gd^{ab0}\pa_{\mu}\phibar^a \lambdabar^b\sigmabar_{\nu}\psi^0
-2gd^{ab0}\phibar^a \lambdabar^b \sigmabar_{\nu}\pa_{\mu}\psi^0\cr
& +2\alpha g_0d^{ab0}\pa_{\mu}\phibar^a \lambdabar^0\sigmabar_{\nu}\psi^b\cr
&+2(1+\alpha)gd^{ab0}\pa_{\mu}\phibar^0 \lambdabar^a\sigmabar_{\nu}\psi^b
\Bigr] \cr}}

The contributions from the graphs shown in Fig.~2 are of the form
\eqn\formtwo{\eqalign{
\sqrt2
&g^3g_CNLC^{\mu\nu}
A_{\mu}^A\phibar^B\lambdabar^C \sigmabar_{\nu}\psi^D
\Bigl(X^{ABCD}_2f^{BAE}d^{CDE}\cr
&+Y^{ABCD}_2f^{DAE}d^{CBE}+Z^{ABCD}_2f^{BDE}d^{CAE}\Bigr)\cr}}
where $g_c\equiv g$. The $X_2$, $Y_2$, $Z_2$ and tensor products in the
decomposition of $X_2^{ABCD}$, $Y_2^{ABCD}$ and $Z_2^{ABCD}$ (as described 
earlier) are shown in Table~2:

\vbox{
\begintable
Fig.| $X_2$| $Y_2$| $Z_2$| Tensor\cr
2a|$\frak12$|$\frak12$|$-\frak12$|$c^Ac^Bc^Cd^D$\cr
2b|$\frak12$|$\frak12$|$\frak12$|$c^Ac^Bc^Cd^D$\cr
2c|$1$|$-1$|$1$|$c^Ac^Bc^Cd^D$\cr
2d|$-1$|$-1$|$-1$|$c^Ac^Bc^Cd^D$\cr
2e|$1$|$0$|$0$|$c^Ac^Bc^Cc^D$\cr
2f|$-\frak14(1-\alpha)$|$\frak14(1-\alpha)$|$-\frak14(1-\alpha)$
|$c^Ac^Bd^Cc^D$\cr
2g|$\frak12$|$\frak12$|$\frak12$|$c^Ac^Bd^Cc^D$\cr
2h|$\frak12\alpha$|$0$|$0$|$c^Ac^B$\cr
2i|$\frak34\alpha$|$0$|$0$|$c^Ac^B$\cr
2j|$-\frak34(3+\alpha)$|$0$|$0$|$c^Ac^B$\cr
2k|$\frak18\alpha$|$-\frak18\alpha$|$\frak18\alpha$|$c^Ac^Bd^Cc^D$\cr
2l|$-\frak38(1-\alpha)$|$-\frak18(1-\alpha)$|$\frak18(1-\alpha)$
|$c^Ac^Bd^Cc^D$\cr
2m|$\frak12\alpha$|$\frak12\alpha$|$-\frak12\alpha$|$c^Ac^Bd^Cc^D$\cr
2n|$\frak12\alpha$|$-\frak12\alpha$|$\frak12\alpha$|$c^Ac^Bc^Cd^D$\cr
2o|$-\frak14\alpha$|$-\frak14\alpha$|$\frak14\alpha$|$c^Ac^Bd^Cc^D$\cr
2p|$\frak38(3+\alpha)$|$-\frak18(3+\alpha)$|$\frak18(3+\alpha)$
|$c^Ac^Bc^Cd^D$\cr
2q|$\alpha$|$0$|$0$|$c^Ac^Bc^Cc^D$\cr
2r|$-\frak14\alpha$|$\frak14\alpha$|$-\frak14\alpha$|$c^Ac^Bc^Cd^D$\cr
2s|$\frak34(1+\alpha)$|$\frak34(1+\alpha)$|$-\frak34(1+\alpha)$|$c^Ac^Bc^Cd^D$\cr
2t|$-\frak12\alpha$|$-\frak12\alpha$|$-\frak12\alpha$|$c^Ac^Bc^Cd^D$\cr
2u|$\frak12\alpha$|$\frak12\alpha$|$\frak12\alpha$|$c^Ac^Bc^Cd^D$\cr
2v|$-\frak38\alpha$|$-\frak38\alpha$|$\frak38\alpha$|$c^Ac^Bc^Cd^D$\cr
2w|$-\frak14(3+\alpha)$|$-\frak14(3+\alpha)$|$-\frak14(3+\alpha)$
|$c^Ac^Bd^Cc^D$\cr
2x|$\frak12$|$\frak12$|$-\frak12$|$c^Ac^Bd^Cc^D$\cr
2y|$1$|$-1$|$1$|$c^Ac^Bd^Cc^D$
\endtable}
\centerline{{\it Table~2:\/} Contributions from Fig.~2}
\vfill
\eject
\vbox{
\begintable
Fig.| $X_2$| $Y_2$| $Z_2$| Tensor\cr
2z|$\frak14(2+\alpha)$|$\frak14(2+\alpha)$|$\frak14(2+\alpha)$
|$c^Ac^Bd^Cc^D$\cr
2aa|$\frak14\alpha$|$\frak14\alpha$|$-\frak14\alpha$|$c^Ac^Bd^Cc^D$\cr
2bb|$\frak34$|$-\frak14$|$\frak14$|$c^Ac^Bd^Cc^D$\cr
2cc|$-\frak14\alpha$|$-\frak14\alpha$|$\frak14\alpha$|$c^Ac^Bd^Cc^D$
\endtable}
\centerline{{\it Table~2:\/} Contributions from Fig.~2 (continued)}
\bigskip

The sum of the contributions from Table~2 can be written in the form
\eqn\sumtwo{\eqalign{
\Gamma_{2\rm{1PI}}^{(1)\rm{pole}}=&\sqrt2
g^3LC^{\mu\nu}A_{\mu}^a\Bigl[
\left(\frak72(1+\alpha)f_{bae}d_{cde}-f_{dae}d_{cbe}+\frak12f_{bde}
d_{cae}\right)Ng
\phibar^b\lambdabar^c \sigmabar_{\nu}\psi^d\cr
&-\frak12(1+5\alpha)\sqrt{2N}g_0
f^{abc}\phibar^b\lambdabar^0 \sigmabar_{\nu}\psi^c
-\frak12(7+5\alpha)\sqrt{2N}gf^{abc}\phibar^b\lambdabar^c \sigmabar_{\nu}\psi^0
\Bigr]\cr
}}

The contributions from Fig.~3 are of the form
\eqn\formthree{
ig^3NLC^{\mu\nu}(\pa_{\mu}A^A_{\nu}
\phibar^B X_3^{ABC}F^C+A^A_{\nu}\pa_{\mu}\phibar^B Y_3^{ABC}F^C)d^{ABC}}
where the $X_3$, $Y_3$ and tensor products in the decomposition of
$X_3^{ABC}$ and $Y_3^{ABC}$ are given in Table 3:
\bigskip
\vbox{
\begintable
Fig.|$X_3$|$Y_3$|Tensor\cr
3a|0|$3$|$c^Ac^Bd^C$\cr
3b|0|$-2$|$c^Ac^Bd^C$\cr
3c|$1$|$1$|$c^Ac^Bd^C$\cr
3d|$-(5+\alpha)$|0|$c^A$\cr
3e|$2\alpha $|$-2$|$c^Ac^Bd^C$   
\endtable}
\centerline{{\it Table~3:\/} Contributions from Fig.~3}

The contributions from Table~3 add to
\eqn\sumthree{\eqalign{
\Gamma_{3\rm{1PI}}^{(1)\rm{pole}}=&
iNg^3LC^{\mu\nu}\Bigl[-(4-\alpha)d^{abc}
\phibar^b \pa_{\mu}A^a_{\nu}F^c\cr
&-3(1-\alpha)d^{ab0}\phibar^a\pa_{\mu}A^b_{\nu}
F^0-
(5+\alpha)d^{ab0}\phibar^0\pa_{\mu}A^a_{\nu}F^b\Bigr].\cr}}

The contributions from Fig.~4 are of the form
\eqn\formfour{
ig^4NLC^{\mu\nu}A^A_{\mu}A^B_{\nu}(X_4^{ABCD}f^{ABE}d^{CDE}
+Y_4^{ABCD}f^{ACE}d^{BDE})\phibar^CF^D}
where the $X_4$ and $Y_4$ and tensor products in the usual decomposition
are given in Table 4:
\bigskip
\vbox{
\begintable
Fig.|$X_4$|$Y_4$|Tensor\cr
4a|$-\frak34\alpha$|$0$|$c^Ac^Bc^Cd^D$\cr
4b|$\frak12\alpha$|$\alpha$|$c^Ac^Bc^Cd^D$\cr
4c|$-\frak12\alpha$|$-\alpha$|$c^Ac^Bc^Cd^D$\cr
4d|$0$|$0$|$c^Ac^Bc^Cd^D$\cr
4e|$\frak14(2+\alpha)$|$2+\alpha$|$c^Ac^Bc^Cd^D$\cr
4f|$-\frak12$|$1$|$c^Ac^Bc^Cd^D$\cr
4g|$-\frak32\alpha$|$0$|$c^Ac^B$\cr
4h|$\frak32(1+\alpha)$|$0$|$c^Ac^B$\cr
4i|$-\frak14(3+\alpha)$|$-(3+\alpha)$|$c^Ac^Bc^Cd^D$\cr
4j|$\frak12\alpha$|$0$|$c^Ac^Bc^Cd^D$\cr
4k|$-\frak34\alpha$|$0$|$c^Ac^Bc^Cd^D$\cr
4l|$0$|$0$|
\endtable}
\centerline{{\it Table~4:\/} Contributions from Fig.~4}

The contributions from Table~4 add to
\eqn\sumfour{\eqalign{
\Gamma_{4\rm{1PI}}^{(1)\rm{pole}}=&ig^4
LC^{\mu\nu}A^a_{\mu}A^b_{\nu}
\Bigl(\frak14(3-4\alpha)Nf^{abe}d^{cde}\phibar^cF^d\cr
&-2\alpha\sqrt{2N}f^{abc}\phibar^cF^0
+\frak32\sqrt{2N}f^{abc}\phibar^0F^c\Bigr).\cr}}

The contributions from Fig.~5 are of the form
\eqn\formfive{
X_5^{ABCD}|C|^2g^2g_Cg_DL\phibar^A\lambdabar^C\lambdabar^DF^B}
where $X_5^{ABCD}$ is given in Table~5.
In Table~5 we have introduced the notation $(\Dtil^A)^{BC}=d^{ABC}$.
Using results from the Appendix, the contributions from Table~5 add to
\eqn\sumfive{\eqalign{
\Gamma_{5\rm{1PI}}^{(1)\rm{pole}}=&
g^4L|C|^2\Bigl[-\frak{1}{2}(3+\alpha)Nf^{ace}f^{bde}
+\frak{11}{8}Nd^{abe}d^{cde}\cr
&-\frak12\delta^{ab}\delta^{cd}-\frak12\delta^{ac}\delta^{bd}\Bigr]
\phibar^a\lambdabar^c\lambdabar^dF^b\cr
&+d^{abc}g^3L|C|^2\sqrt{2N}\Bigl[g\phibar^0\lambdabar^a\lambdabar^bF^c
+3g_0\phibar^a\lambdabar^b\lambdabar^0F^c
+2g\phibar^a\lambdabar^b\lambdabar^cF^0\Bigr]\cr
&+4g^3g_0L|C|^2\left(\phibar^0\lambdabar^0\lambdabar^aF^a
+2\phibar^a\lambdabar^a\lambdabar^0F^0\right).\cr}}
\bigskip
\vbox{
\begintable
Fig.|$X_5^{ABCD}$\cr
5a|0\cr
5b|$4\tr[\Ftil^A\Ftil^C\Dtil^B\Dtil^D]$\cr
5c|$-2\tr[\Ftil^A\Dtil^C\Ftil^D\Dtil^B]$\cr
5d|$-\alpha Nd^Cc^Dc^Xd^{ABX}d^{CDX}$\cr
5e|$(1+\alpha)Nc^Xd^{ABX}d^{CDX}$\cr
5f|$-\frak12N\alpha c^Ad^Bc^Xd^{ABX}d^{CDX}$\cr
5g|$0$\cr
5h|$2\alpha \tr[\Ftil^C\Ftil^A\Dtil^B\Dtil^D]$\cr
5i|$-\frak12(3+\alpha)\tr[\Ftil^A\Ftil^B\Ftil^D\Ftil^C]$\cr
5j|$\frak12\alpha (\tr[\Ftil^C\Ftil^A\Ftil^D\Ftil^B]-\frak12
Nf^{XAC}f^{XBD})$\cr
5k|$(\tr[\Ftil^A\Ftil^C\Ftil^D\Ftil^B]+\frak12
Nf^{XAC}f^{XBD})$
\endtable}
\centerline{{\it Table~5:\/} Contributions from Fig.~5}

The divergent contributions
to the effective action from the graphs in Fig. 6 are of the form
\eqn\formsixa{
iLNg^3X_6C^{\alpha\beta}d^{abe}f^{cde}\phibar^a
\phibar^b\psi^c_{\alpha}\psi^d_{\beta}}
where the contributions from the
individual graphs to $X_6$ and the associated tensors in the usual
decomposition are given in Table 6:
\bigskip
\vbox{
\begintable
Fig.|$X_6$\cr
6a|$0$\cr
6b|$0$\cr
6c|$\alpha$\cr
6d|$-\alpha$\cr
6e|$-1$
\endtable}
\centerline{{\it Table~6:\/} Contributions from Fig.~6}
\bigskip

The contributions from Table~6 add to
\eqn\sumsixa{
\Gamma_{6\rm{1PI}}^{(1)\rm{pole}}
=-iLNg^3C^{\alpha\beta}d^{abe}f^{cde}\phibar^a
\phibar^b\psi^c_{\alpha}\psi^d_{\beta}.}

The divergent contributions
to the effective action from the graphs in Fig. 7 are of the form
\eqn\formsix{
imLNg^2g_AX^{ABC}_7C^{\mu\nu}d^{ABC}\pa_{\mu}A_{\nu}^A\phibar^B
\phibar^C}
where the contributions from the
individual graphs to $X_7$ and the associated tensors in the usual 
decomposition are given in Table 7:
\bigskip
\vbox{
\begintable
Fig.|$X_7$|Tensor\cr
7a|$2$|$c^Ac^Bd^C$\cr
7b|$-1$|$c^Ac^Bd^C$\cr
7c|$-1$|$c^Ac^Bd^C$\cr
7d|$0$|\cr
7e|$-4$|$d^Ac^Bc^C$\cr
7f|$-2\alpha$|$d^Ac^Bc^C$\cr
7g|$-2$|$d^Ac^Bc^C$\cr
7h|$-\frak12$|$c^Ac^Bd^C$\cr
7i|$-1$|$c^Ac^Bd^C$\cr
7j|$-(1+2\alpha)$|$c^Ac^Bd^C$\cr
7k|$\frak32$|$c^Ac^Bd^C$\cr
7l|$-\frak12(5+\alpha)$|$c^A$\cr
7m|$\alpha$|$d^Ac^Bc^C$\cr
7n|$1$|$d^Ac^Bc^C$\cr
7o|$\frak12$|$c^Ac^Bd^C$\cr
7p|$1$|$c^Ac^Bd^C$\cr
7q|$1+2\alpha$|$c^Ac^Bd^C$\cr
7r|$-\frak32$|$c^Ac^Bd^C$
\endtable}
\centerline{{\it Table~7:\/} Contributions from Fig.~7}
\bigskip
These results add to
\eqn\sumsix{\eqalign{
\Gamma_{7\rm{1PI}}^{(1)\rm{pole}}=
&-\frak12(5+\alpha)iLg^2C^{\mu\nu}m\Bigl[3Ngd^{abc}\pa_{\mu}A_{\nu}^a\phibar^b 
\phibar^c\cr
&+2g\sqrt{2N}\pa_{\mu}A_{\nu}^a\phibar^a\phibar^0
+4g_0\sqrt{2N}\pa_{\mu}A_{\nu}^0\phibar^a\phibar^a\Bigr].\cr}}
(Note that the contributions from Figs.~7(h-k) cancel those from 
Figs.~7(o-r) respectively.)

The divergent contributions
to the effective action from the graphs in Fig. 8 are of the form
\eqn\formseven{
imLNg^4X_8^{ABCD}C^{\mu\nu}f^{ABE}d^{CDE}A_{\mu}^AA_{\nu}^B\phibar^C
\phibar^D}
where the contributions from the
individual graphs to $X_8$ and the associated tensors in the usual 
decomposition are given in Table 8:
\bigskip
\vbox{
\begintable
Fig.|$X_8$|Tensor\cr
8a|$-2$|$c^Ac^Bc^Cc^D$\cr
8b|$1$|$c^Ac^Bc^Cc^D$\cr
8c|$1$|$c^Ac^Bc^Cc^D$\cr
8d|$2$|$c^Ac^Bc^Cc^D$\cr
8e|$\alpha$|$c^Ac^Bc^Cc^D$\cr
8f|$1$|$c^Ac^Bc^Cc^D$\cr
8g|$-\frak14(3+\alpha)$|$c^Ac^Bc^Cd^D$\cr
8h|$0$|\cr
8i|$0$|\cr
8j|$1$|$c^Ac^Bc^Cd^D$\cr
8k|$\frak34\alpha$|$c^Ac^Bc^Cd^D$\cr
8l|$-\frak12\alpha$|$c^Ac^Bc^Cd^D$\cr
8m|$\frak34\alpha$|$c^Ac^Bc^Cd^D$\cr
8n|$\frak14(2+\alpha)$|$c^Ac^Bc^Cd^D$\cr
8o|$0$|\cr
8p|$0$|\cr
8q|$-\frak34\alpha$|$c^Ac^B$\cr
8r|$\frak34(1+\alpha)$|$c^Ac^B$\cr
8s|$-\frak12\alpha$|$c^Ac^Bc^Cc^D$\cr
8t|$-\frak12$|$c^Ac^Bc^Cc^D$
\endtable}
\centerline{{\it Table~8:\/} Contributions from Fig.~8}
\bigskip  
\vbox{
\begintable
Fig.|$X_8$|Tensor\cr
8u|$\frak14(3+\alpha)$|$c^Ac^Bc^Cd^D$\cr
8v|$0$|\cr
8w|$0$|\cr
8x|$-1$|$c^Ac^Bc^Cd^D$\cr
8y|$-\frak34\alpha$|$c^Ac^Bc^Cd^D$\cr
8z|$\frak12\alpha$|$c^Ac^Bc^Cd^D$\cr
8aa|$-\frak34\alpha$|$c^Ac^Bc^Cd^D$\cr
8bb|$-\frak14(2+\alpha)$|$c^Ac^Bc^Cd^D$\cr
8cc|$0$|\cr
8dd|$0$|
\endtable}
\centerline{{\it Table~8:\/} Contributions from Fig.~8 (continued)}
\bigskip
These results add to
\eqn\sumseven{
\Gamma_{8\rm{1PI}}^{(1)\rm{pole}}=
iLg^4C^{\mu\nu}mf^{abe}A_{\mu}^aA_{\nu}^b\Bigl[
\frak14(13+2\alpha)Nd^{cde}\phibar^c 
\phibar^d+\frak32\sqrt{2N}\phibar^e\phibar^0\Bigr]}
(Note that the contributions from Figs.~8(g-p) cancel those from
Figs.~8(u-dd) respectively.)

The contributions from Fig.~9 are of the form
\eqn\formeight{
X_9^{ABCD}g^2g_Cg_DmL|C|^2\phibar^A\phibar^B\lambdabar^C\lambdabar^D.}
The contributions from the individual graphs to 
$X_9^{ABCD}$ are given in Table~9.
The results in Table~9 add to
\eqn\sumeight{\eqalign{
\Gamma_{9\rm{1PI}}^{(1)\rm{pole}}=&
|C|^2mL\Bigl\{\bigl[\frak{11}{8}Nd^{abe}d^{cde}
-\frak12\left(1-4\frak{g^2}{g_0^2}
\right)\delta^{ab}\delta^{cd}
-\frak12\delta^{ad}\delta^{bc}\cr
&+\frak4N\left(1-\frak{g^2}{g_0^2}\right)f^{ace}f^{bde}\bigr]
g^4\phibar^a\phibar^b\lambdabar^c\lambdabar^d\cr
&+g^3d^{abc}\sqrt{2N}\phibar^a\lambdabar^b\left(3g_0\phibar^c\lambdabar^0
+g\phibar^0\lambdabar^c\right)
+4g^3g_0\phibar^a\phibar^0\lambdabar^a\lambdabar^0\Bigr\}.\cr}}
(Note that the contributions from Figs.~9(h--m) cancel those from 
Figs.~9(u--z); this is analogous to the situation with
Figs.~7 and 8, and is a consequence of our choice of coefficient for
the last term in Eq.~\lagmass.) 
\bigskip
\vbox{
\begintable   
Fig.|$X_9^{ABCD}$\cr
9a|$\frak12\alpha \tr[\Ftil^A\Ftil^B\Dtil^C\Dtil^D]$\cr
9b|$\frak12\tr[\Ftil^A\Ftil^B\Dtil^C\Dtil^D]$\cr
9c|$\frak12(3+\alpha)\tr[\Ftil^A\Ftil^B\Dtil^C\Dtil^D]$\cr
9d|$-\alpha \tr[\Ftil^A\Ftil^B\Dtil^C\Dtil^D]$\cr
9e|$Nd^{ABE}d^{CDE}c^Ac^Bc^Cc^Dc^E-2c^Cc^D\tr[\Ftil^A\Ftil^B\Dtil^C\Dtil^D]
+\frak4Nf^{ACE}f^{BDE}$
\crnorule
|$+\frak{2g^2}{g_0^2}\left(c^Ac^Bc^Cc^D\delta^{AB}\delta^{CD}
-\frak2Nf^{ACE}f^{BDE}\right)$\cr
9f|$\frak12\alpha \tr[\Ftil^A\Ftil^B\Ftil^C\Ftil^D]$\cr
9g|$\frak12\tr[\Ftil^A\Ftil^B\Ftil^C\Ftil^D]$\cr
9h|$-\tr[\Ftil^A\Ftil^B\Ftil^C\Ftil^D]-\frak12Nf^{ACE}f^{BDE}$\cr
9i|$-\frak12\alpha \left(\tr[\Ftil^A\Ftil^C\Ftil^B\Ftil^D]
-\frak12Nf^{ACE}f^{BDE}\right)$\cr
9j|$-2\alpha\tr[\Ftil^C\Ftil^A\Dtil^B\Dtil^D]$\cr
9k|$-4\tr[\Ftil^C\Ftil^A\Dtil^D\Dtil^B]$\cr
9l|$\frak12\alpha Nc^Ad^Bc^Ed^{ABE}d^{CDE}$\cr
9m|$2\tr[\Ftil^A\Dtil^C\Ftil^D\Dtil^B]$\cr
9n|$0$\cr
9o|$0$\cr
9p|$-\frak12\alpha Nd^{ABE}d^{CDE}d^Cc^Dc^E$\cr
9q|$\frak12(1+\alpha)Nd^{ABE}d^{CDE}c^E$\cr
9r|$-\frak14(3+\alpha)\tr[\Ftil^A\Ftil^B\Ftil^C\Ftil^D]$\cr
9s|$-\frak14\alpha \tr[\Ftil^A\Ftil^B\Ftil^C\Ftil^D]$\cr
9t|$-\frak14\tr[\Ftil^A\Ftil^B\Ftil^C\Ftil^D]$\cr
9u|$\tr[\Ftil^A\Ftil^B\Ftil^C\Ftil^D]+\frak12Nf^{ACE}f^{BDE}$\cr
9v|$\frak12\alpha\left(\tr[\Ftil^A\Ftil^C\Ftil^B\Ftil^D]
-\frak12Nf^{ACE}f^{BDE}\right)$\cr
9w|$2\alpha\tr[\Ftil^C\Ftil^A\Dtil^B\Dtil^D]$\cr
9x|$4\tr[\Ftil^C\Ftil^A\Dtil^D\Dtil^B]$
\endtable}
\centerline{{\it Table~9:\/} Contributions from Fig.~9}
\vfill
\eject
\vbox{
\begintable
Fig.|$X_9^{ABCD}$\cr
9y|$-\frak12\alpha Nc^Ad^Bc^Ed^{ABE}d^{CDE}$\cr
9z|$-2\tr[\Ftil^A\Dtil^C\Ftil^D\Dtil^B]$\cr
9aa|$0$\cr
9bb|$0$
\endtable}
\centerline{{\it Table~9:\/} Contributions from Fig.~9 (continued)}
\bigskip
The results from Fig.~10 are of the form
\eqn\formnine{
Ny g^2LX_{10}C^{\mu\nu}f^{abc}\pa_{\mu}\phibar^a\pa_{\nu}\phibar^b\phibar^c}
and the contributions from the individual graphs to $X$ are given in 
Table~10.
\bigskip  
\vbox{
\begintable
Fig.|$X_{10}$\cr
10a|$\frak12\alpha$\cr
10b|$0$\cr
10c|$2$\cr
10d|$0$\cr
10e|$-\frak32$
\endtable}
\centerline{{\it Table~10:\/} Contributions from Fig.~10}
\bigskip   
The results in Table~10 add to
\eqn\sumnine{\eqalign{
\Gamma_{10\rm{1PI}}^{(1)\rm{pole}}=&
\frak12 Ny g^2LC^{\mu\nu}(1+\alpha)
f^{abc}\pa_{\mu}\phibar^a\pa_{\nu}\phibar^b\phibar^c
\cr}}
We have not explicitly drawn most of 
the diagrams (labelled Fig.~(11a,b$\ldots$)) giving contributions of the form
\eqn\formten{
iC^{\mu\nu}y g^2g_D
L\left(X_{11}^{ABCD}\pa_{\mu}\phibar^A\phibar^B\phibar^CA_{\nu}^D
+Y_{11}^{ABCD}\phibar^A\phibar^B\phibar^C\pa_{\mu}A_{\nu}^D\right),}
since they can be obtained by adding external scalar lines to the diagrams 
of Fig.~7. Thus Figs.~11(e-o) are obtained from Figs.~7(a-k) by adding an 
external scalar ($\phibar$) line at the position of the
cross. Figs.~11(p-v) are obtained from Figs.~7(l-r) by adding an 
external scalar ($\phibar$) line at the position of the dot. The remaining
Figs.~11(a-d) are depicted in Fig.~11. The 
individual contributions to $X_{11}^{ABCD}$ and 
$Y_{11}^{ABCD}$ in Eq.~\formten\
are given in Table~11.
\bigskip
\vbox{
\begintable
Fig.|$X_{11}^{ABCD}$|$Y_{11}^{ABCD}$\cr
11a|$-\alpha\tr[\Ftil^A\Ftil^B\Ftil^C\Ftil^D]$|$0$\cr
11b|$-\tr[\Ftil^A\Ftil^B\Ftil^C\Ftil^D]$|$0$\cr
11c|$\frak12\alpha(\tr[\Ftil^A\Ftil^B\Ftil^C\Ftil^D]-\frak12
Nf^{ABE}f^{CDE})$|$0$\cr
11d|$0$|$0$\cr
11e|$-4\tr[\Dtil^B\Dtil^A\Ftil^C\Ftil^D]$|$0$\cr
11f|$2\tr[\Dtil^A\Dtil^B\Ftil^C\Ftil^D]$|$0$\cr
11g|$2\tr[\Dtil^A\Ftil^B\Dtil^C\Ftil^D]$|$0$\cr
11h|$-2Nf^{ABE}f^{CDE}$|$0$\cr
11i|$0$|$-4\tr[\Dtil^B\Ftil^C\Dtil^D\Ftil^A]$\cr
11j|$0$|$-2\alpha\tr[\Ftil^B\Ftil^C\Dtil^D\Dtil^A]$\cr
11k|$0$|$-2\tr[\Ftil^B\Ftil^C\Dtil^D\Dtil^A]$\cr
11l|$\tr[\Dtil^{(A}\Dtil^{B)}\Ftil^C\Ftil^D]$|$0$\cr
11m|$2\tr[\Ftil^D\Ftil^A\Dtil^B\Dtil^C]$|$0$\cr
11n|$2\tr[\Dtil^B\Dtil^C\Ftil^D\Ftil^A]$|$-2\alpha\tr[\Dtil^B\Dtil^C\Ftil^D\Ftil^A]$\cr
11o|$-3\tr[\Dtil^B\Dtil^C\Ftil^D\Ftil^A]$|$0$\cr
11p|$0$|$\frak16(5+\alpha)\left(\tr[\Ftil^A\Ftil^B\Ftil^C\Ftil^D]
-Nc^Dd^{ABE}d^{CDE}\right)$\cr
11q|$\frak23\alpha\left(\tr[\Ftil^{[A}\Ftil^{B]}\Ftil^C\Ftil^D]
-\frak12Nf^{ABE}f^{CDE}\right)$|
$\frak13\alpha\left(4\tr[\Ftil^A\Ftil^B\Dtil^C\Dtil^D]
+Nc^Bc^Cd^Ed^{ADE}d^{BCE}\right)$\cr
11r|$\tr[\Ftil^A\Ftil^B\Ftil^C\Ftil^D]$
|$\frak13\tr[\Ftil^A\Ftil^B\Ftil^C\Ftil^D]$\crnorule
|$$|$+\frak13\left(4\tr[\Ftil^A\Ftil^B\Dtil^C\Dtil^D]
+Nc^Bc^Cd^Ed^{ADE}d^{BCE}\right)$\cr
11s|$\frak13\left((3+\alpha)Nf^{ABE}f^{CDE}-\tr[\Ftil^A\Ftil^B\Ftil^C\Ftil^D]
\right)$
|$\frak16(1+\alpha)\tr[\Ftil^A\Ftil^B\Ftil^C\Ftil^D]$\crnorule
|$+\frak16\left(4\tr[\Ftil^A\Ftil^D\Dtil^B\Dtil^C]+c^Ac^Dd^ENd^{ADE}d^{BCE}\right)$
|$+\frak16\left(4\tr[\Ftil^A\Ftil^D\Dtil^B\Dtil^C]+c^Ac^Dd^ENd^{ADE}d^{BCE}\right)$\cr
11t|$\frak16\left(4\tr[\Ftil^A\Ftil^B\Ftil^C\Ftil^D]-\alpha Nf^{ABE}f^{CDE}
\right)$|$-\frak13\alpha\tr[\Ftil^A\Ftil^B\Ftil^C\Ftil^D]$\crnorule
|$-\frak13\left(4\tr[\Ftil^D\Ftil^A\Dtil^B\Dtil^C]
+Nc^Ac^Dd^Ed^{ADE}d^{BCE}\right)$|
\endtable}
\centerline{{\it Table~11:\/} Contributions from Fig.~11}
\bigskip
\vbox{
\begintable
Fig.|$X_{11}^{ABCD}$|$Y_{11}^{ABCD}$\cr
11u|$\tr[\frak16\left\{(4+3\alpha)\Ftil^A\Ftil^B-
3\alpha\Ftil^B\Ftil^A\right\}\Ftil^C\Ftil^D]
$|$-\frak16\alpha\tr[\Ftil^A\Ftil^B\Ftil^C\Ftil^D]$\crnorule
|$-\frak14\alpha Nf^{ABE}f^{CDE}$|
$+\frak13\alpha
\left(4\tr[\Ftil^D\Ftil^A\Dtil^B\Dtil^C]+Nc^Ac^Dd^Ed^{ADE}d^{BCE}\right)$\crnorule
|$-\frak13\left(4\tr[\Ftil^D\Ftil^A\Dtil^B\Dtil^C]+Nc^Ac^Dd^Ed^{ADE}d^{BCE}\right)$|\cr
11v|$-\tr[\Ftil^A\Ftil^B\Ftil^C\Ftil^D]+\frak32Nf^{ABE}f^{CDE}$|$0$\crnorule
|$+\frak12\left(4\tr[\Ftil^D\Ftil^A\Dtil^B\Dtil^C]+Nc^Ac^Dd^Ed^{ADE}d^{BCE}\right)$|
\endtable}
\centerline{{\it Table~11:\/} Contributions from Fig.~11 (continued)}
\bigskip
The results sum to
\eqn\sumten{\eqalign{
\Gamma_{11\rm{1PI}}^{(1)\rm{pole}}=&iC^{\mu\nu}y g^2L\Bigl(
-\frak12g\left(3+\frak73\alpha\right)Nf^{abe}f^{cde}\pa_{\mu}\phibar^a
\phibar^b\phibar^cA_{\nu}^d\cr
&+\left[-\left(\frak54-\frak16\alpha\right)Nd^{abe}d^{cde}
+\left(3+\frak73\alpha\right)\delta^{ab}\delta^{cd}\right]
g\phibar^a\phibar^b\phibar^c\pa_{\mu}A_{\nu}^d\cr
&-\frak12(9+\alpha)g\sqrt{2N}d^{abc}\phibar^0\phibar^a\phibar^b
\pa_{\mu}A_{\nu}^c-(5+\alpha)g\phibar^0\phibar^0\phibar^a 
\pa_{\mu}A_{\nu}^a\cr
&-2g_0\sqrt{2N}d^{abc}\phibar^a\phibar^b\phibar^c\pa_{\mu}A_{\nu}^0
-8g_0\phibar^a\phibar^a\phibar^0\pa_{\mu}A_{\nu}^0\Bigr).\cr
}}
\appendix{B} {Group identities for $U(N)$}
The basic
commutation relations for $U(N)$ are (for the fundamental representation):
\eqn\commrel{ [R^a,R^b]=if^{abc}R^c,\quad
\{R^A,R^B\}=d^{ABC}R^C,}
where $d^{ABC}$ is totally symmetric.
Defining matrices $\Ftil^A$, $\Dtil^A$ by $(\Ftil^A)^{BC}=if^{BAC}$, 
$(\Dtil^A)^{BC}=d^{ABC}$,
useful identities for $U(N)$ are
\eqn\sunidents{\eqalign{
\Tr[\Ftil^A\Ftil^B]=&N\delta^{AB},\qquad \Tr[\Dtil^A\Dtil^B]=N\delta^{AB},\cr
\Tr[\Ftil^A\Ftil^B\Dtil^C]=&\frak{N}{2}d^{ABC}c^Ac^Bd^C,\qquad
\Tr[\Ftil^A\Dtil^B\Dtil^C]=i\frak{N}{2}f^{ABC},\cr
f^{ABE}d^{CDE}+&f^{ACE}d^{DBE}+f^{ADE}d^{BCE}=0,\cr
f^{ABE}f^{CDE}=&d^{ACE}d^{BDE}-d^{ADE}d^{BCE},\cr}}
and also
\eqn\sunidentsa{\eqalign{
\Tr[\Ftil^A\Ftil^B\Ftil^C\Ftil^D]=&c^Ac^Bc^Cc^D\Bigl[
\frak12\delta^{(AB}\delta^{CD)}\cr
&+\frak{N}{4}\left(d^{ABE}d^{CDE}+d^{ADE}d^{BCE}-d^{ACE}d^{BDE}\right)\Bigr],\cr
\Tr[\Ftil^A\Ftil^B\Ftil^C\Dtil^D]=&-\frak{N}{4}i(d^{ABE}f^{CDE}+f^{ABE}d^{CDE})
c^Ac^Bc^Cd^D,\cr
\Tr[\Ftil^A\Ftil^B\Dtil^C\Dtil^D]=&\Bigl[\frak12c^Ac^Bc^Cc^D\left(
\delta^{AB}\delta^{CD}-\delta^{AC}\delta^{BD}-\delta^{AD}\delta^{BC}\right)\cr
&+\frak{N}{4}c^Ac^Bd^Cd^D\left(
d^{ABE}d^{CDE}+d^{ADE}d^{BCE}-d^{ACE}d^{BDE}\right)\Bigr],\cr
\Tr[\Ftil^A\Dtil^B\Ftil^C\Dtil^D]=&
c^Ac^Bc^Cc^D\frak12\left(\delta^{AC}\delta^{BD}
-\delta^{AB}\delta^{CD}-\delta^{AD}\delta^{BC}\right)\cr
&+\frak{N}{4}c^Ad^Bc^Cd^D\left(
d^{ABE}d^{CDE}+d^{ADE}d^{BCE}-d^{ACE}d^{BDE}\right)\Bigr].\cr
}}
\vfill
\eject
\epsfysize= 5in
\centerline{\epsfbox{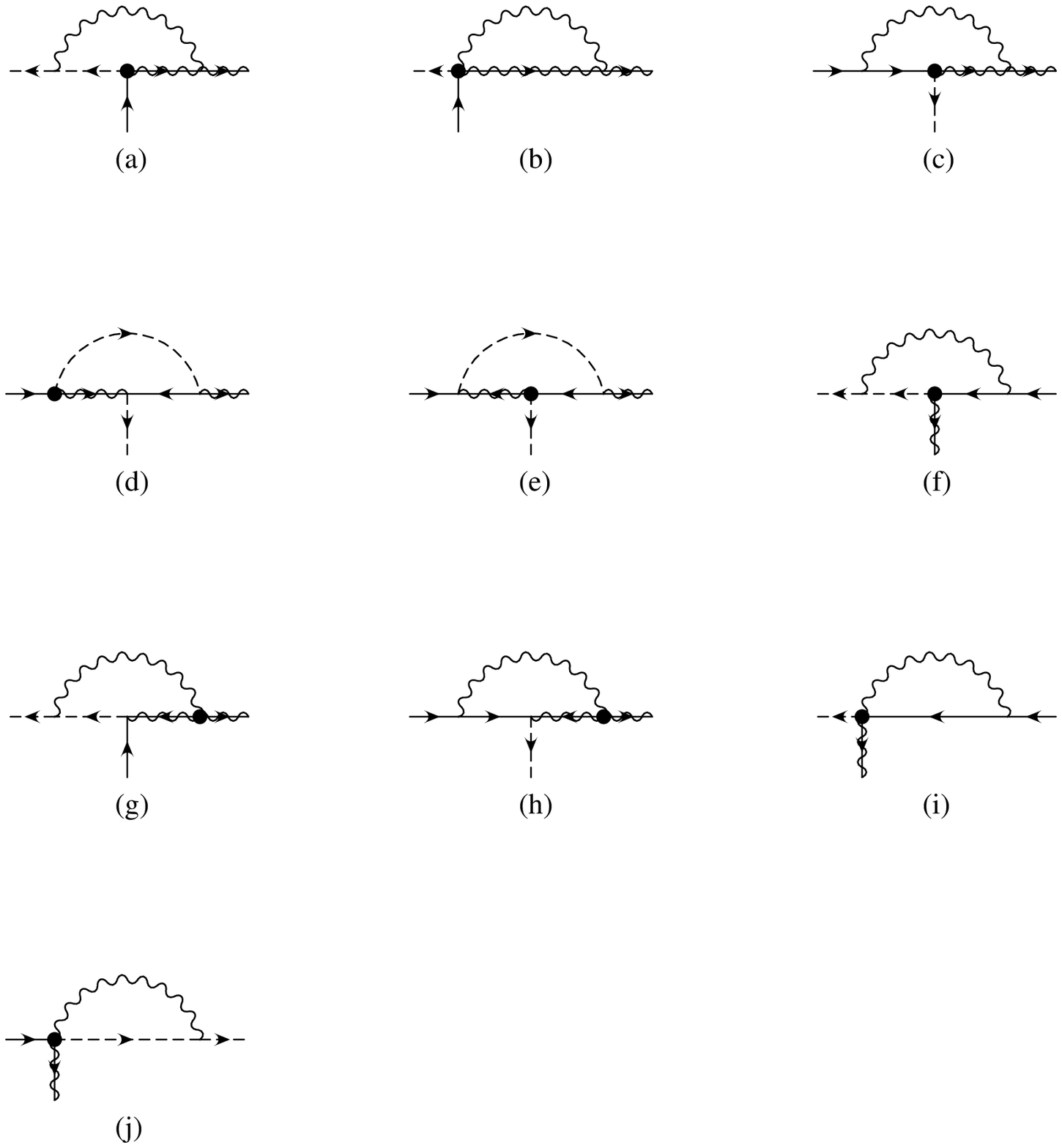}}
\inparg
{\it \noindent Fig. 1: Diagrams with one gaugino, one scalar and one
chiral fermion line; the dot represents the position of a $C$.} 
\medskip
\outparg
\vfill  
\eject  
\epsfysize= 7in
\centerline{\epsfbox{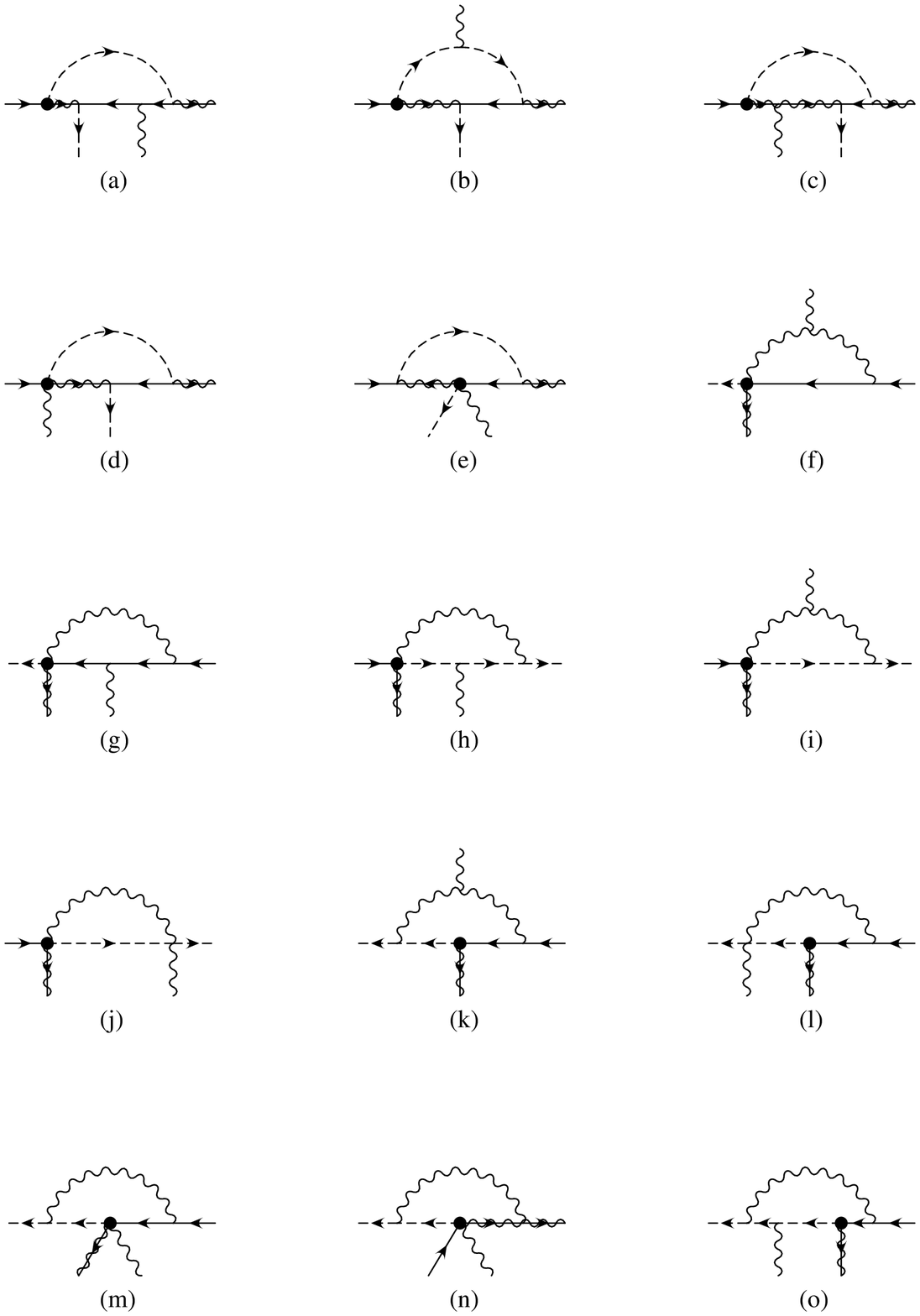}}
\inparg
{\it \noindent Fig. 2: Diagrams with one gaugino, one scalar, one
chiral fermion and one gauge line; the dot represents the position of a $C$.}
\medskip
\outparg

\bigskip
\epsfysize= 7in
\centerline{\epsfbox{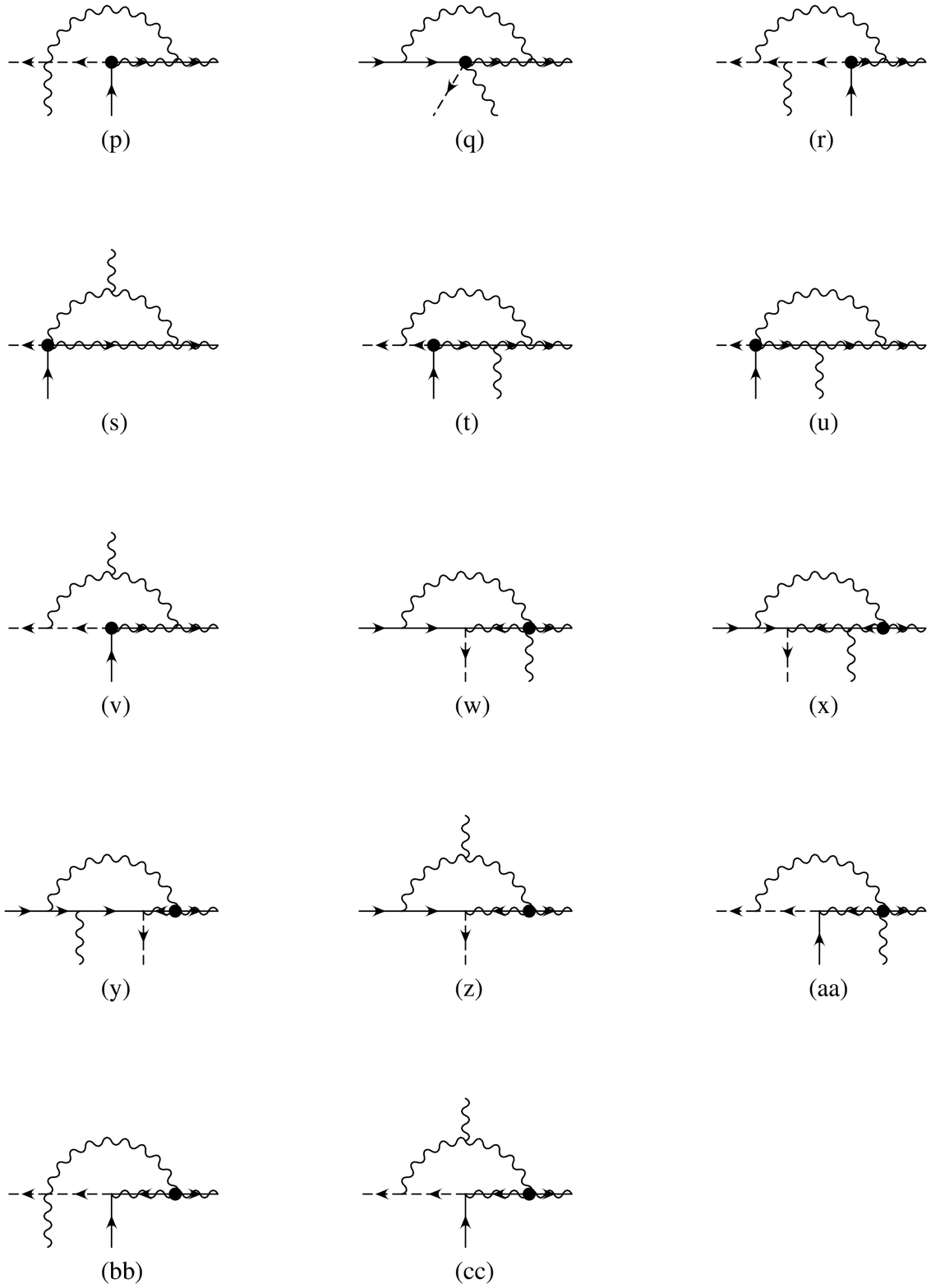}}
\inparg
{\it \noindent Fig. 2(continued).}   
\medskip
\outparg

\epsfysize= 2.5in
\centerline{\epsfbox{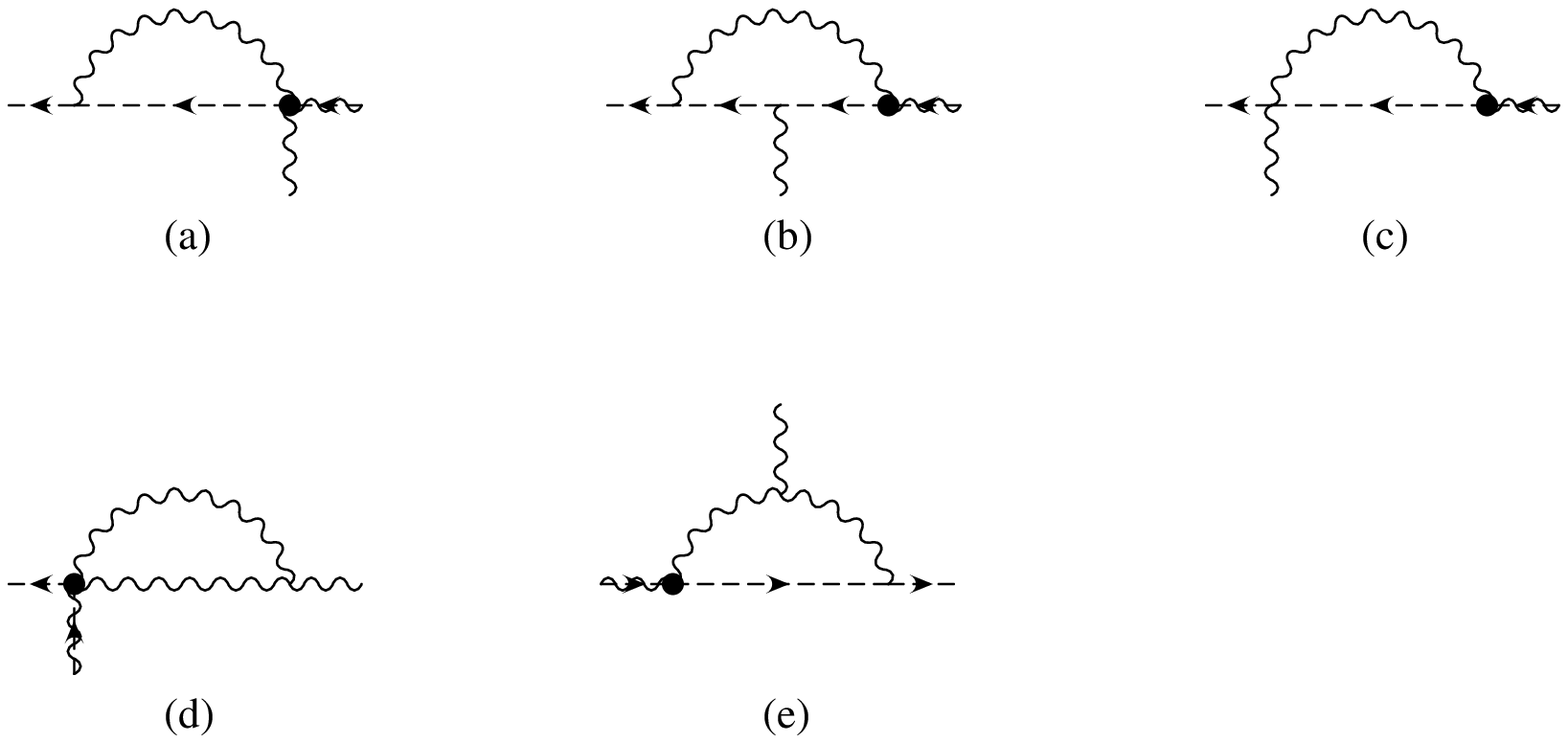}}
\inparg 
{\it \noindent Fig. 3: Diagrams with one gauge, one scalar and one
auxiliary line; the dot represents the position of a $C$.}
\medskip
\outparg
\vfill  
\eject  
\epsfysize= 5in
\centerline{\epsfbox{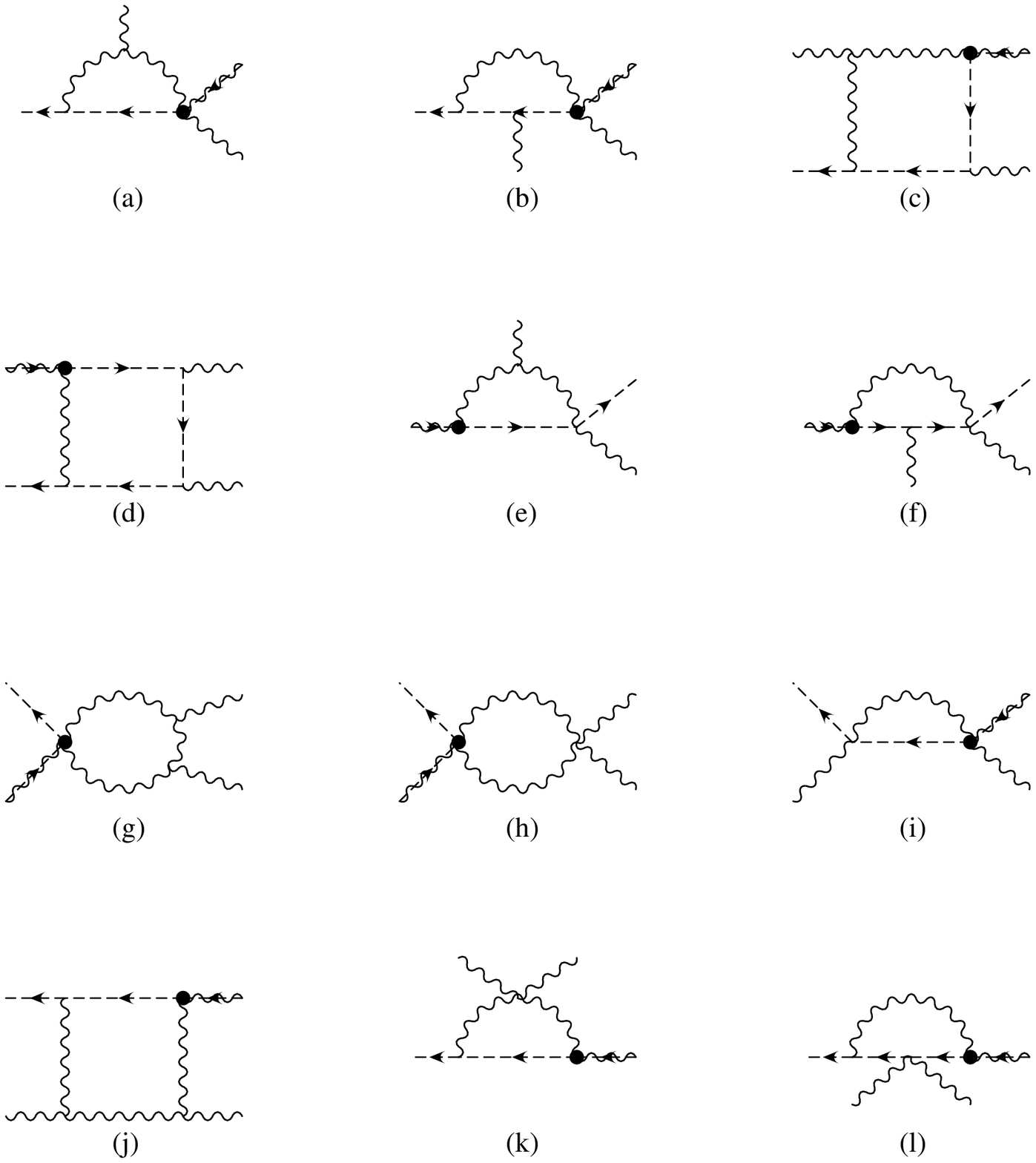}}
\inparg
{\it \noindent Fig. 4: Diagrams with two gauge, one scalar and one
auxiliary line; the dot represents the position of a $C$.}
\medskip
\outparg
\vfill  
\eject  
\bigskip
\epsfysize= 5in
\centerline{\epsfbox{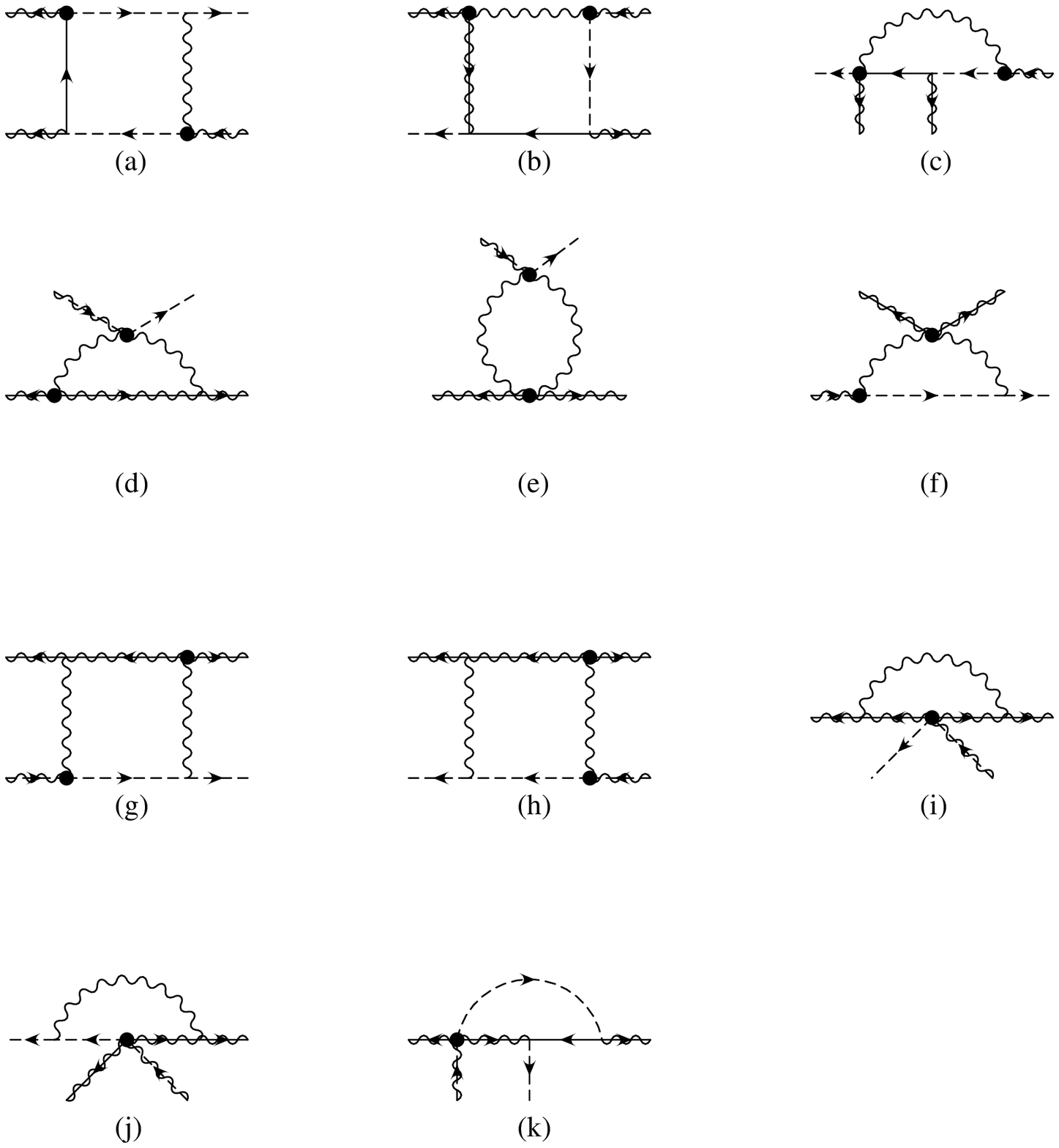}}
\inparg
{\it \noindent Fig. 5: Diagrams with two gaugino, one scalar and one
auxiliary line; the dot represents the position of a $C$ or a $|C|^2$.}
\medskip
\outparg
\vfill
\eject

\bigskip
\epsfysize= 2.5in
\centerline{\epsfbox{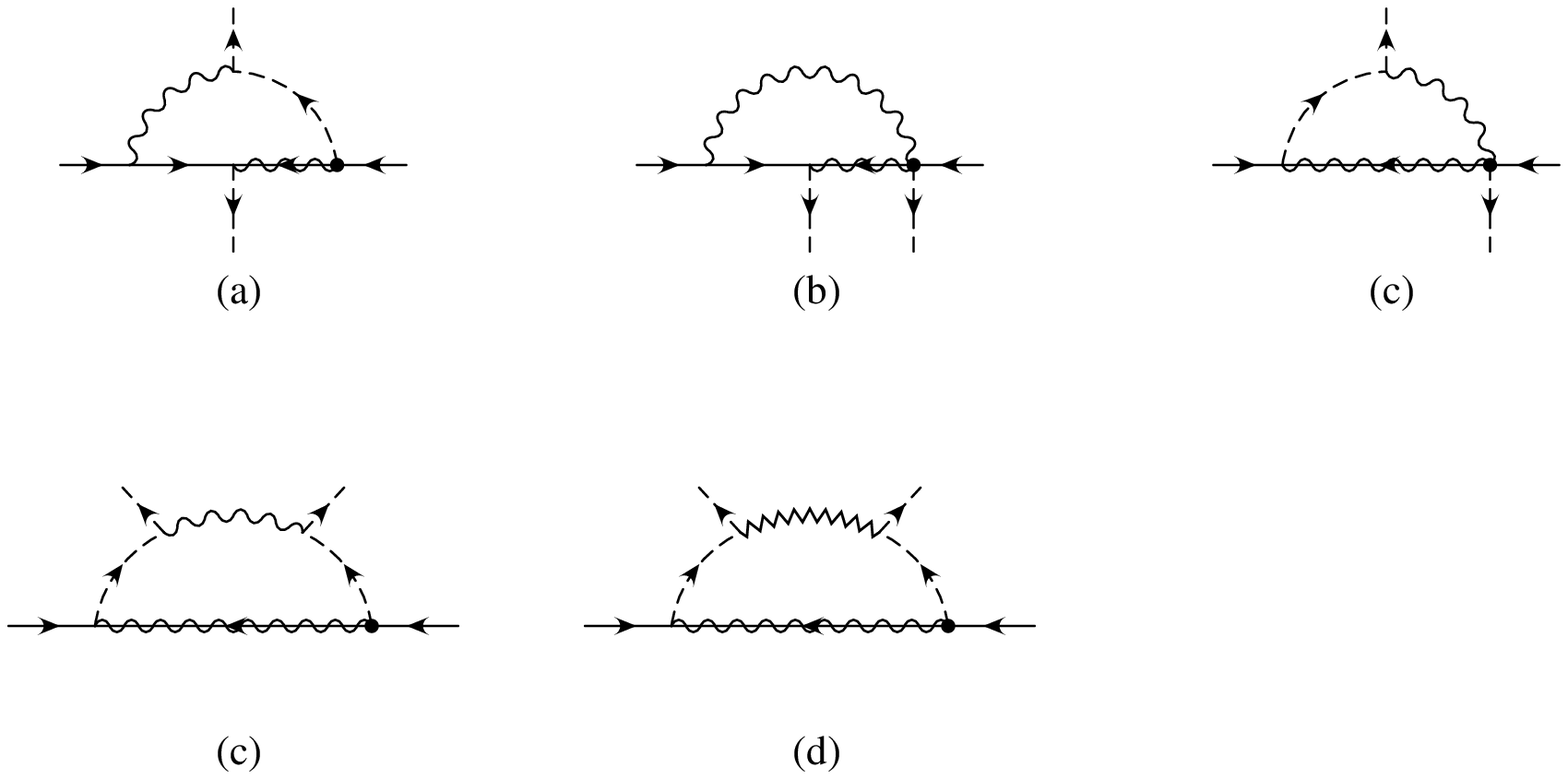}}
\inparg
{\it \noindent Fig. 6: Diagrams with two scalar
and two chiral fermion lines; the dot represents the position of a $C$.}
\medskip
\outparg

\vfill  
\eject  
\epsfysize= 7in
\centerline{\epsfbox{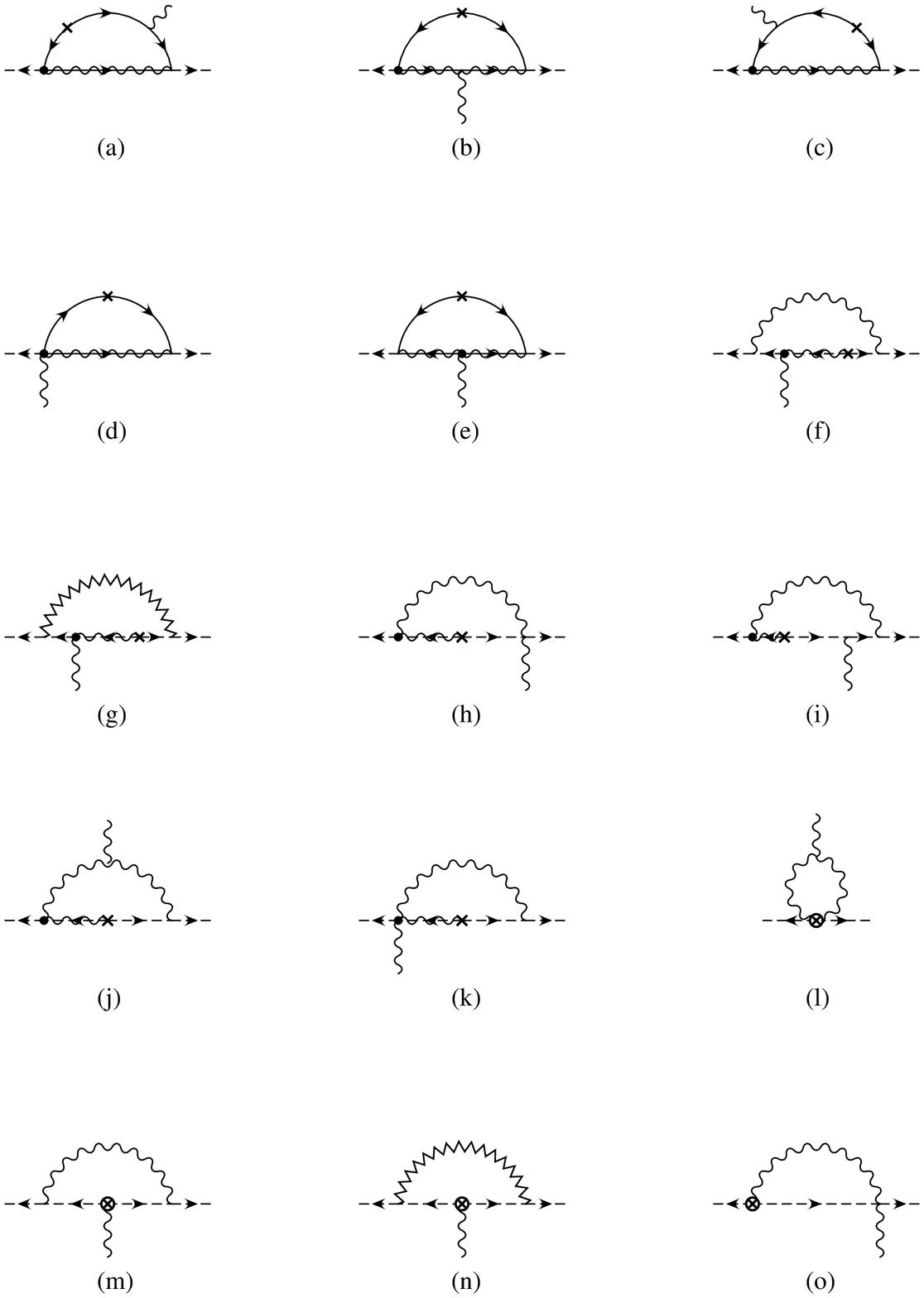}}
\inparg
{\it \noindent Fig. 7: Diagrams with two scalar,
one gauge line; a dot denotes a $C$, a cross a mass
and a crossed circle a vertex with both a mass and a $C$.}
\medskip
\outparg
\bigskip
\epsfysize= 1.2in
\centerline{\epsfbox{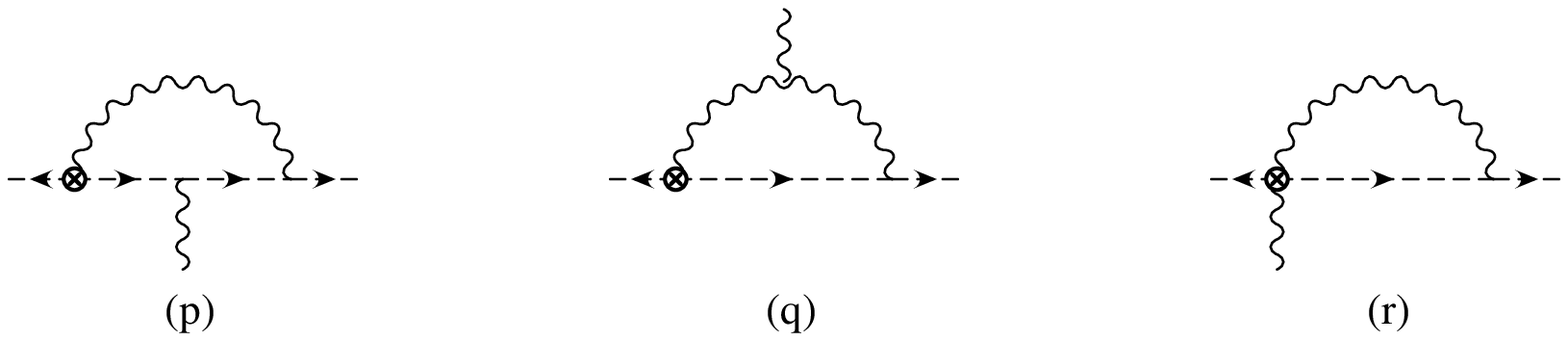}}
\inparg
{\it \noindent Fig. 7(continued)}   
\medskip
\outparg

\vfill
\eject 
\epsfysize= 7in
\centerline{\epsfbox{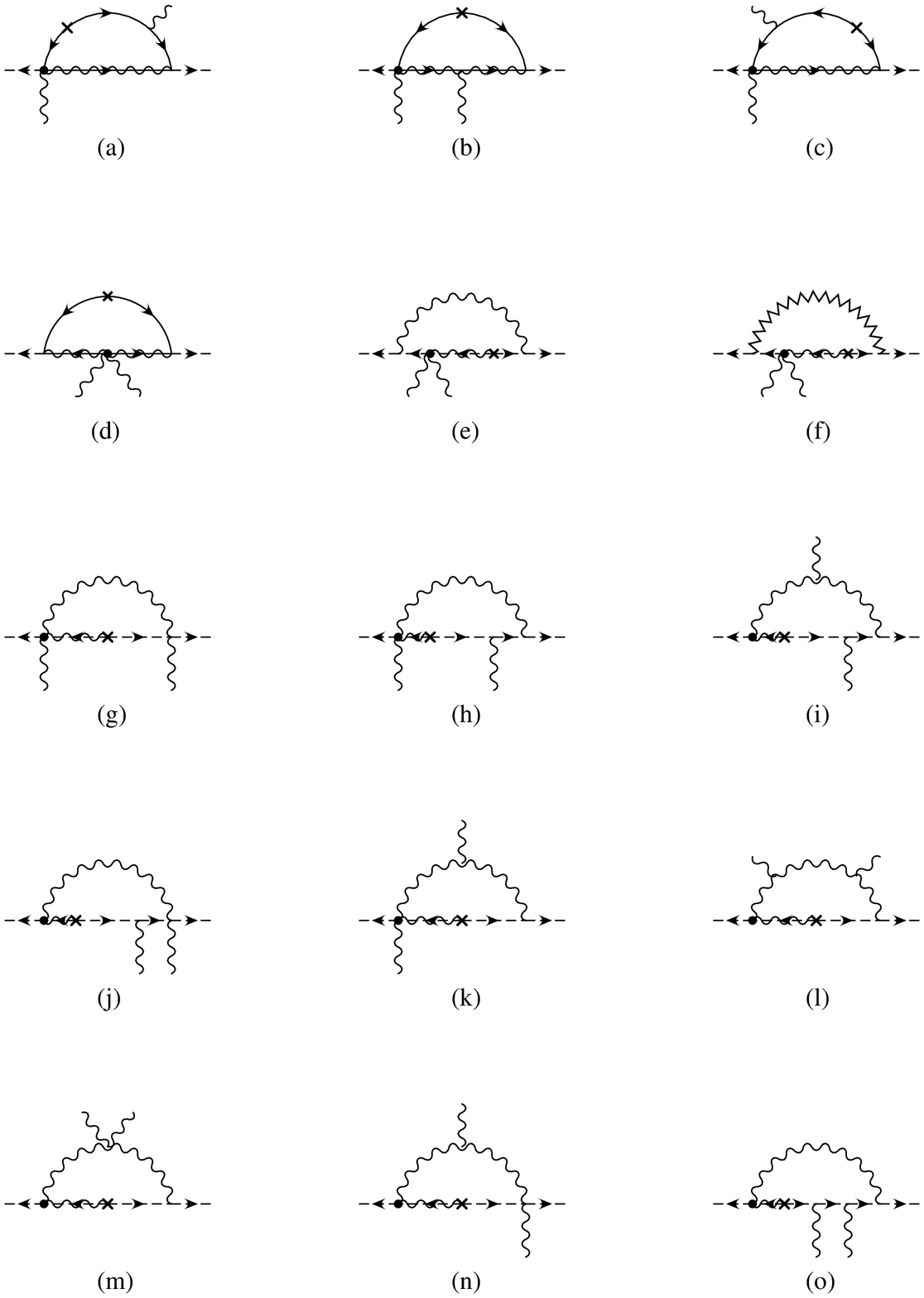}}
\inparg 
{\it \noindent Fig. 8: Diagrams with two scalar, two
gauge lines; a dot denotes a $C$, a cross a mass
and a crossed circle a vertex with both a mass and a $C$.}
\medskip
\outparg
\vfill
\eject 
\epsfysize= 7in
\centerline{\epsfbox{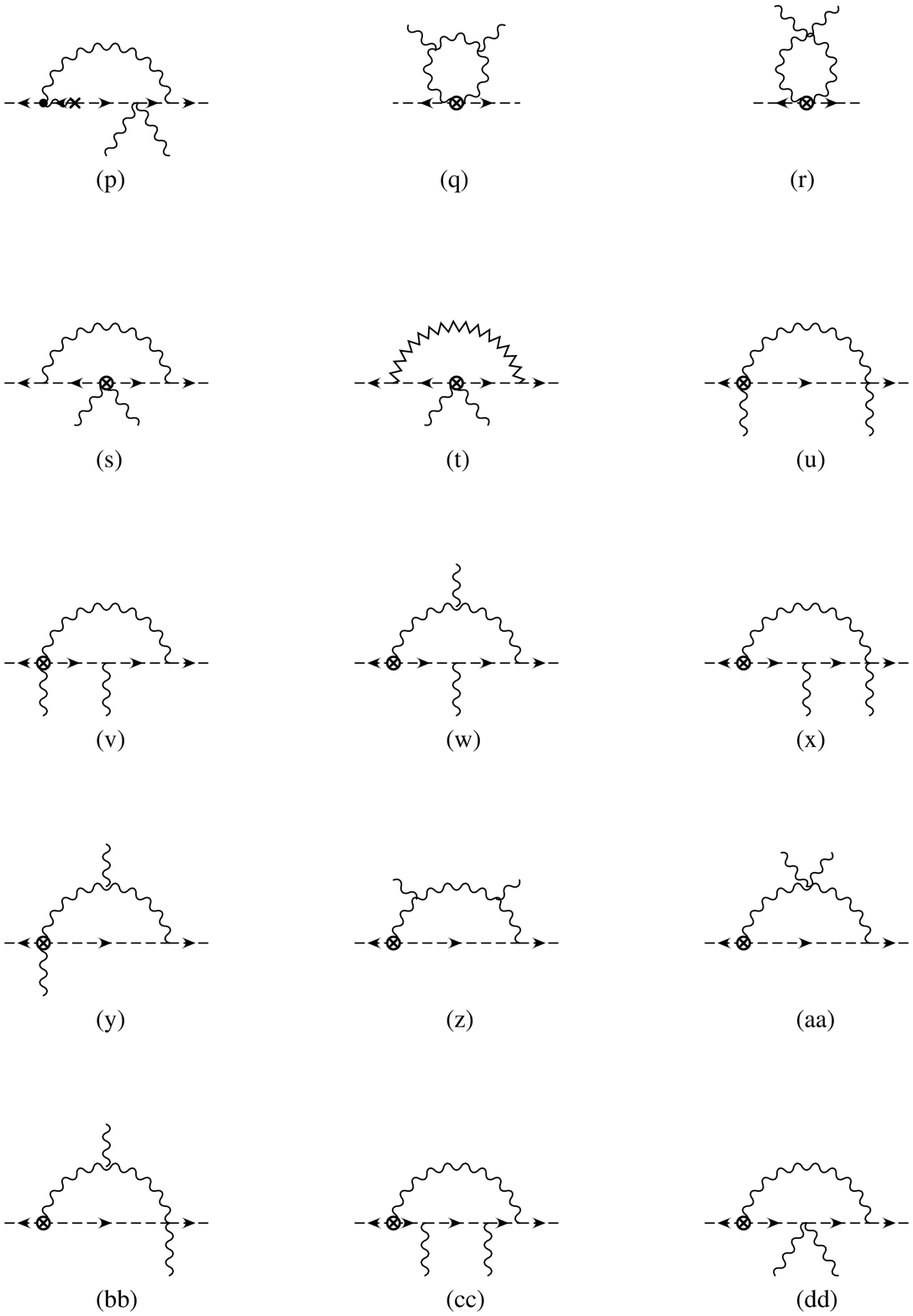}}
\inparg 
{\it \noindent Fig. 8(continued)}
\medskip
\outparg

\vfill
\eject  
\epsfysize= 7in
\centerline{\epsfbox{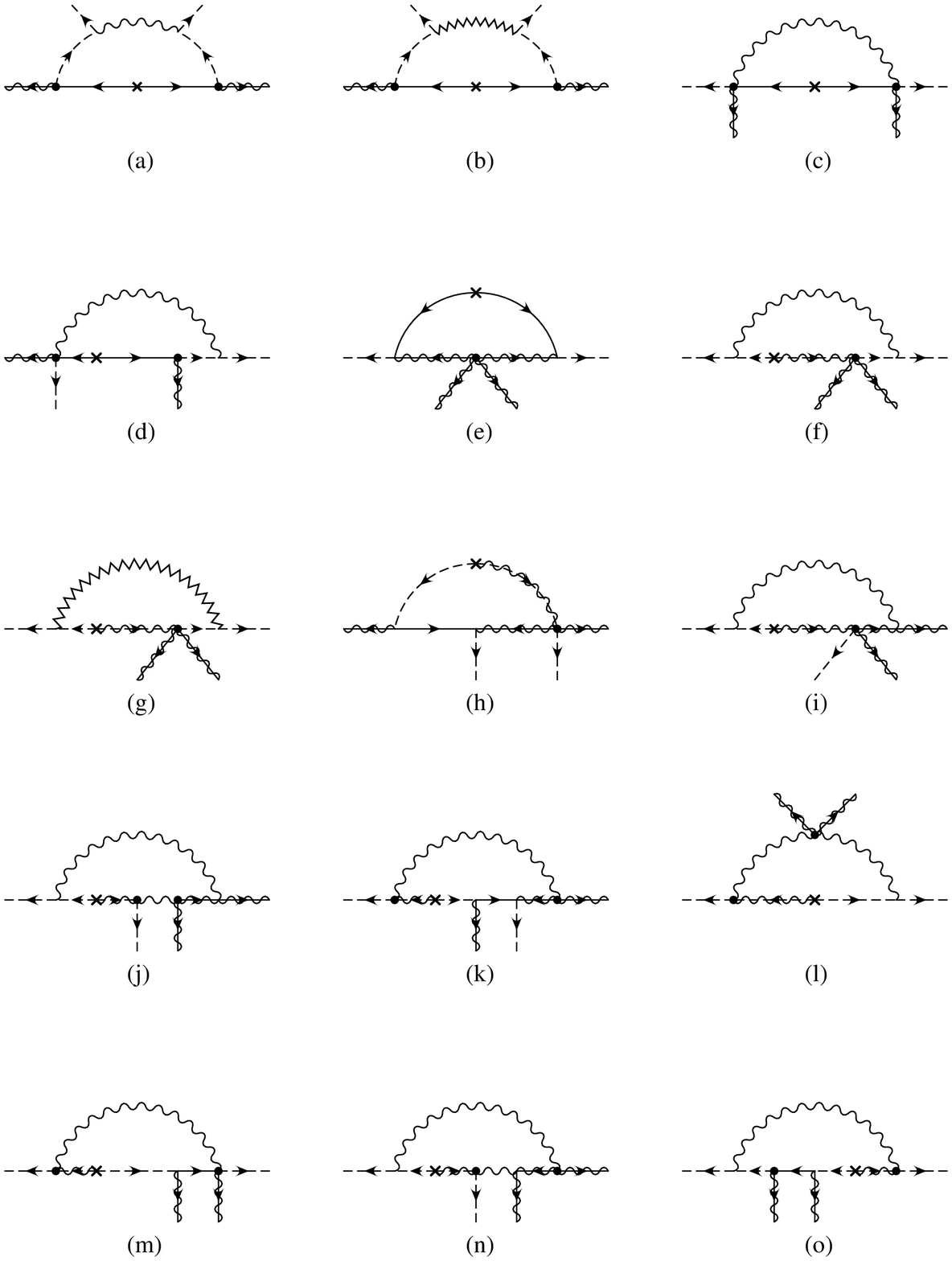}}
\inparg 
{\it \noindent Fig. 9: Diagrams with two scalar, two gaugino lines; 
a dot denotes a $C$, a cross a mass  
and a crossed circle a vertex with both a mass and a $C$.}
\medskip
\outparg
\vfill 
\eject
\epsfysize= 7in
\centerline{\epsfbox{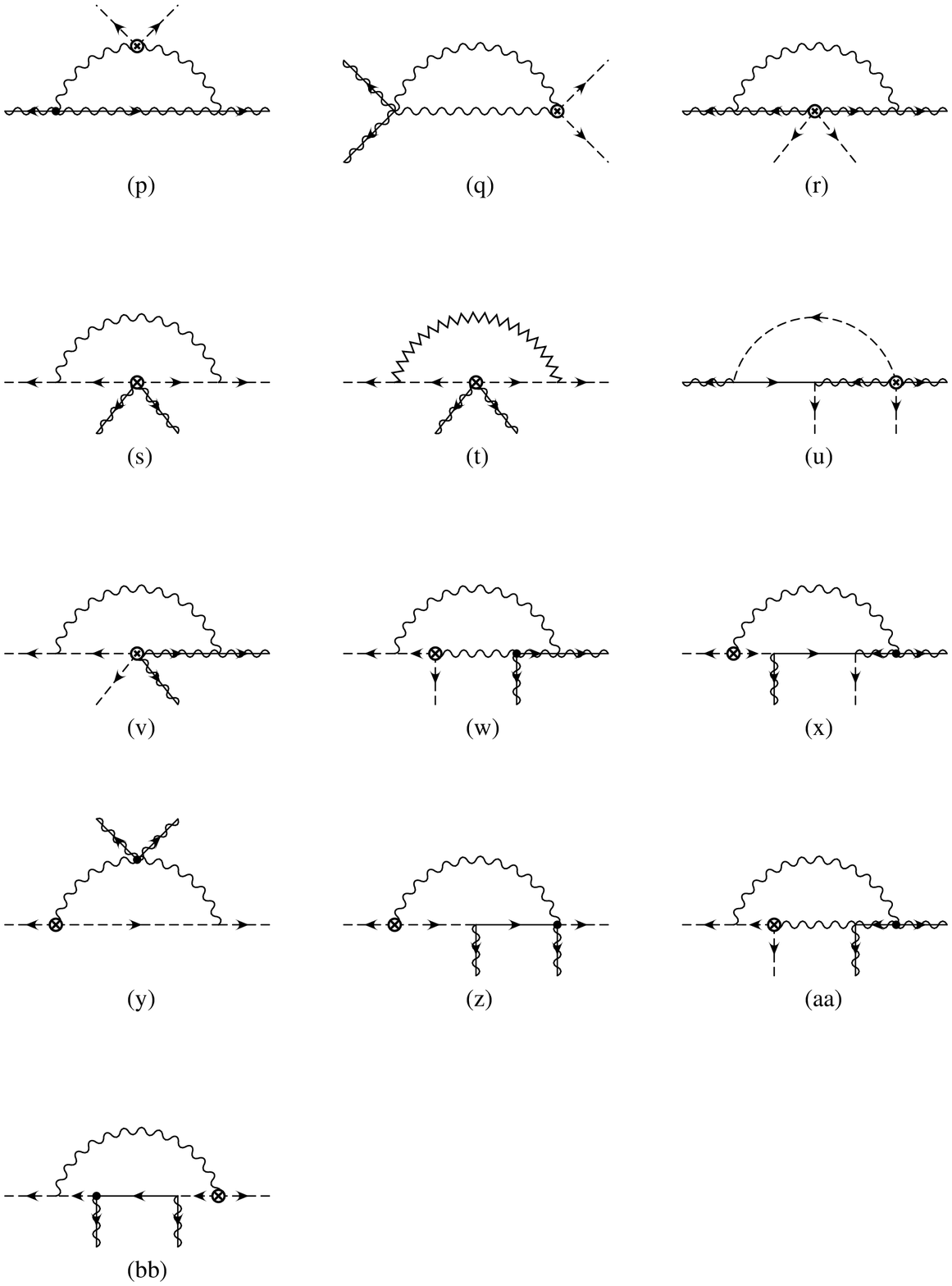}}
\inparg
{\it \noindent Fig. 9 (continued)}
\medskip
\outparg
\vfill
\eject
\epsfysize= 2.5in
\centerline{\epsfbox{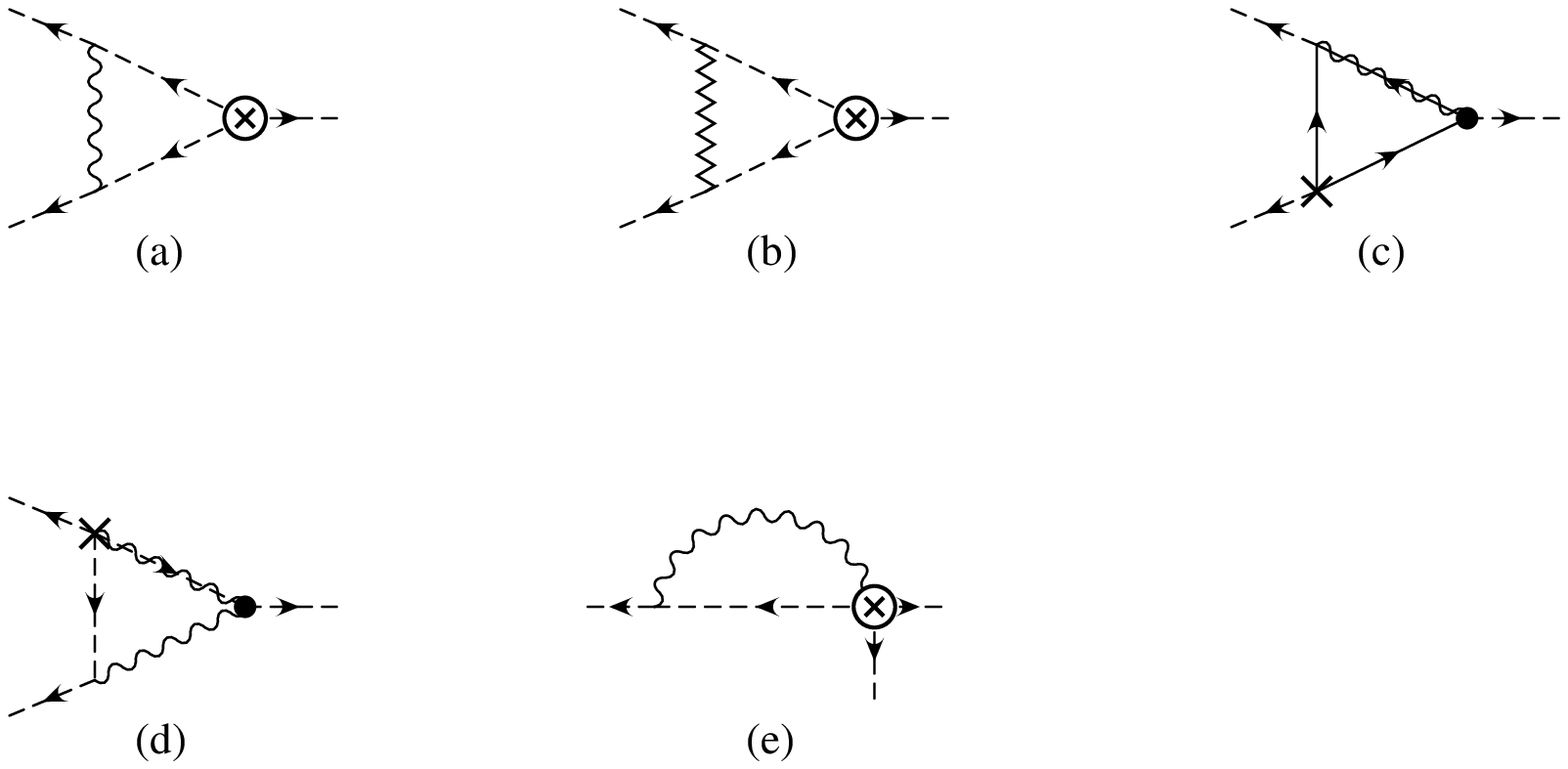}}
\inparg 
{\it \noindent Fig. 10: Diagrams with three scalar lines; a dot represents 
the position of a $C$, a cross a superpotential vertex without a $C$
and a crossed circle a superpotential vertex with a $C$.}
\medskip
\outparg

\bigskip
\epsfysize= 2.5in
\centerline{\epsfbox{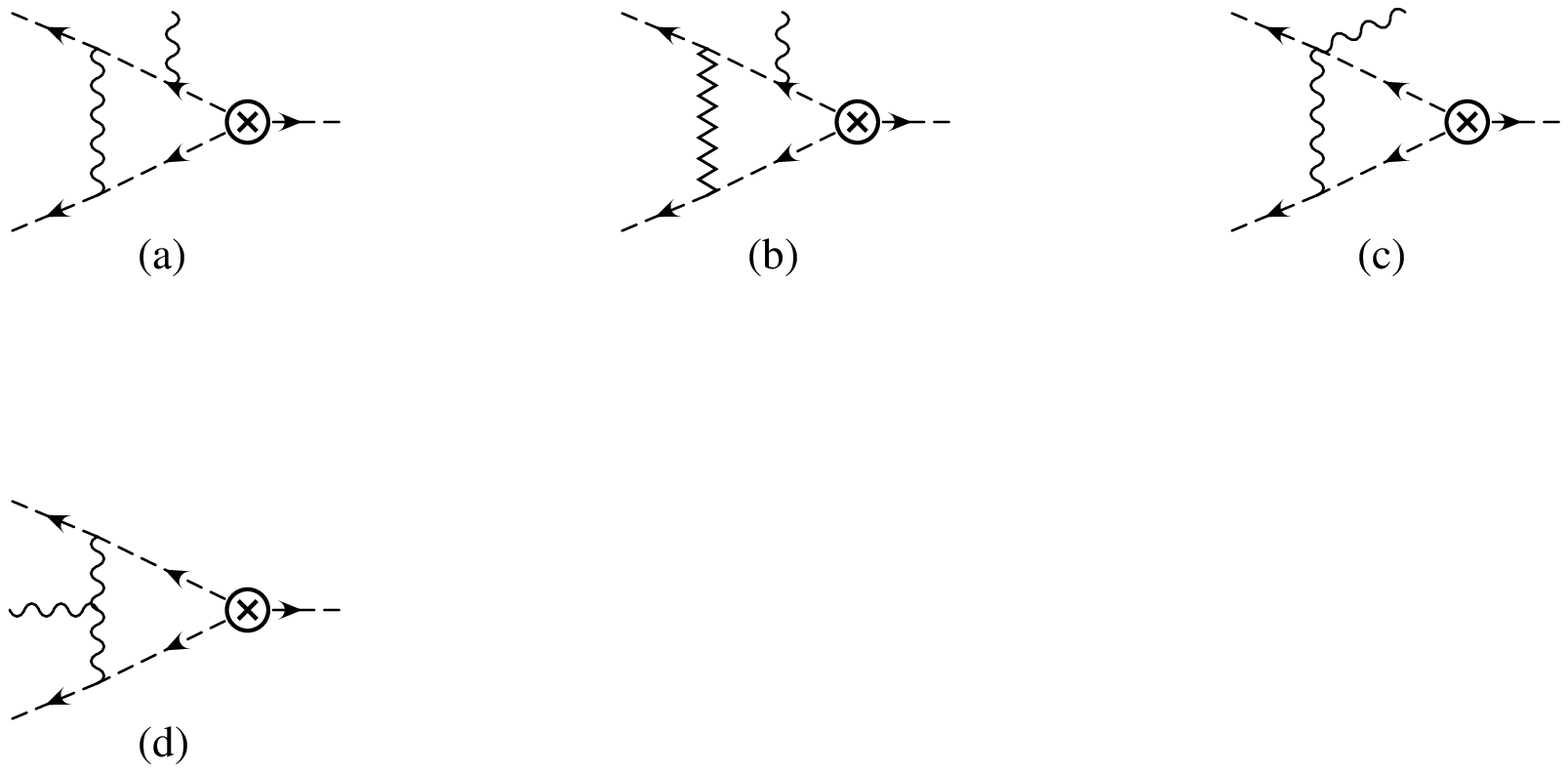}}
\inparg
{\it \noindent Fig. 11: Diagrams with three scalar lines and one gauge
line; a crossed circle represents a superpotential vertex with a $C$.}
\medskip
\outparg

\listrefs
\bye